\documentclass[11pt,a4paper]{article}
\usepackage[sort]{cite}
\usepackage{url,colortbl}
\usepackage{xcolor}
\usepackage{bm,amsmath,amssymb}
\usepackage[]{graphicx}
\usepackage[normalem]{ulem}
\usepackage{bbding}
\usepackage{enumitem}
\usepackage{duckuments}
\usepackage[margin=2.5cm]{geometry}
\graphicspath{{./Images/}}
\usepackage{authblk}
\usepackage{xr}
\usepackage{hyperref}

\usepackage{mathtools}
\usepackage{ulem}

\newcommand{\dd}{\text{d}}

\newcommand{\nI}{n_\text{I}}
\newcommand{\nN}{n_\text{N}}
\newcommand{\nS}{n_\text{S}}

\begin{document}

\title{Optimal control in combination therapy for heterogeneous cell populations with drug synergies}
\author[1]{Simon F. Martina-Perez\thanks{Email: simon.martina-perez@medschool.ox.ac.uk}}
\author[2]{Samuel W.S. Johnson}
\author[2]{Rebecca M. Crossley}
\author[3]{Jennifer C. Kasemeier}
\author[3]{Paul M. Kulesa}
\author[2]{Ruth E. Baker}
\affil[1]{School of Medicine and Biomedical Sciences, University of Oxford, Oxford, UK}
\affil[2]{Mathematical Institute, University of Oxford, Oxford, UK}
\affil[3]{Children's Mercy Hospital Research Institute, Kansas City, MO, USA}

\date{}
\maketitle
\begin{abstract}
    Cell heterogeneity plays an important role in patient responses to drug treatments.
    In many cancers, it is associated with poor treatment outcomes.
    Manymodern drug combination therapies aim to exploit cell heterogeneity, but determining how to optimise responses from heterogeneous cell populations while accounting for multi-drug synergies remains a challenge. 
    In this work, we introduce and analyse a general optimal control framework that can be used to model the treatment response of multiple cell populations that are treated with multiple drugs that mutually interact.
    In this framework, we model the effect of multiple drugs on the cell populations using a system of coupled semi-linear ordinary differential equations and derive general results for the optimal solutions. 
    We then apply this framework to three canonical examples and discuss the wider question of how to relate mathematical optimality to clinically observable outcomes, introducing a systematic approach to propose qualitatively different classes of drug dosing inspired by optimal control. 
\end{abstract}

\section{Introduction}
Combination drug therapies are a mainstay of modern treatments for a wide range of diseases with cancer perhaps being the most notable example \cite{lieftink2021takes, Marjanovich13, Altschuler10, Hausser20, Crucitta22, Proietto23}. The success of combination therapies relative to single-drug therapies stems from their ability to exploit drug synergies \cite{lieftink2021takes}, target a diverse cell population within a target tissue \cite{Marjanovich13, Altschuler10, Hausser20, Crucitta22}, and avoid the toxicity associated with high doses of a single drug \cite{lieftink2021takes}. However, designing combination treatments is not straightforward, owing to the difficulties in predicting not only the responses of different cells present in the target tissues to any individual drug, but also responses to all possible drug combinations. Making such predictions is all the more important in settings where the presence of multiple cell types -- or the presence of cells with different features, such as where they are in the cell cycle -- is clinically significant. For example, cell type heterogeneity in cancer is associated with poor prognosis, which may be due to sub-populations of cancer cells that are more proliferative, or less sensitive to drug treatment \cite{Dagogo18,Crucitta22,Proietto23, Gomez22, Hausser20}. In other diseases, such as type I diabetes \cite{Benninger22} and hepatic fibrosis \cite{Ramachandran20}, diversity of the target cell populations is also known to be clinically important. Therefore, to design combination treatments, one needs to carefully describe how each cell type will respond over time to the drug combinations present in a putative treatment. At the same time, one needs to take into account practical considerations, such as drug toxicity and cost. Optimal control theory provides an established framework to combine differential equation modelling of the tissue response to treatment with a cost function that models treatment effect, as well as additional penalties for factors such as cost and toxicity \cite{lenhart07optimalcontrol}. However, a general ODE framework for multi-drug actions on different discrete cell populations has not yet been developed. In this paper, we propose such a framework, and explicitly compute optimal control solutions that correspond to various treatment scenarios. 

The benefit of developing a general model for multi-drug, multi-cell population interactions lies in its applicability to a wide range of problems. Previously, ODE models have been used to predict the effect of drug treatments on heterogeneous cell populations \cite{lenhart07optimalcontrol, Pillai23, Clairambault23, Kashkooli20, Kirschner96, Wang16}. However, the large amount of (phenomenological) functional forms employed in such models reduces their transferability between problems and hinders the ability to systematically analyse the resulting optimal controls. In this work we show how a general model can capture important cell-drug interactions while easily identifying optimal pharmacodynamic regimes. This can, in turn, guide novel insights into designing therapies. The framework introduced in this paper captures three key phenomena to first order. First, we model cell proliferation and death, assuming a linear growth rate in the absence of drugs, which is appropriate at small population numbers and few environmental factors limiting growth. Such growth can be mediated by drugs: for example, paclitaxel is a chemotherapeutic agent \cite{markman2002paclitaxel} that prevents mitosis. Secondly, we allow cells to spontaneously convert to another cell type at fixed rates \cite{Yuan16, Marjanovich13}, which may also be mediated by drug treatment. For instance, in neuroblastoma, undifferentiated sympathoblast cells can convert to neuroblasts in the presence of retinoic acid (RA) \cite{Gomez22, Zeineldin22awry, Jansky21Origin}. Finally, several drugs can interact with one another to produce synergistic effects \cite{lieftink2021takes}. Hence, drug-mediated cell differentiation and proliferation pathways should be made to depend on combinations of drugs in the model. 

Moreover, the model presented in this work describes the direct influence of the effective drug concentrations present in the relevant target tissues, \textit{i.e.} we consider an control problem for the optimal pharmacodynamics. In so doing, we explicitly ignore the pharmacokinetics, which describe how the administered drugs are absorbed, metabolised, and cleared from the system. This is a necessary choice for a generalised framework, since the pharmacokinetics of different drugs can vary vastly, for example, depending on the drugs' metabolic properties, or route of administration (such as oral, intramuscular or intravenous). Hence, pharmacokinetics are an inherently application-specific problem. While understanding the specific dosing that needs to be administered in time to target tissues is key to implement therapies downstream, this work will limit itself with modelling what effective drug concentrations should be present in the target tissues, regardless of how they are administered. Subsequent efforts can then be made to couple a tailored, application-specific model corresponding to the specifics of individual drugs and the minutiae of administration.

The structure of this work is as follows. In Section~\ref{section:modelling} we introduce the general modelling framework. In Section~\ref{section:optimalControl} we compute the optimality conditions and the optimal control policy arising from our framework. We then apply this framework to three canonical examples for treating heterogeneous (cancer) cell populations. In Section~\ref{section:example1} we provide a first example to control a two-population model, based on ovarian cancer cells treated with a synergistic combination of chemotherapy drugs \cite{lieftink2021takes}. In Section~\ref{section:example2} we explore an optimal control formulation for controlling cell populations in neuroblastoma through drugs targeting the tropomyosin A and B signalling pathways. Finally in Section~\ref{section:example3} we introduce a way to extend our framework to control the relative abundances of cell types within a population. Together, our work provides a template model that is widely applicable for the control of heterogeneous cell populations and their responses to treatment with several, interacting drugs. Example code has been made available for all computations at our GitHub repository: \url{https://github.com/SWSJChCh/multiplicativeControl}.

\section{Modelling treatment response of a heterogeneous cell population with multiplicative control and drug-drug interaction}
\label{section:modelling}
In this section, we introduce a general ODE model for the treatment response of a heterogeneous cell population with drug synergies. In Section~\ref{section:GeneralFormulation} we formulate the general model and in Section~\ref{section:FunctionalForm} we propose a simple functional form for the model.

\subsection{ODE model: general formulation}
\label{section:GeneralFormulation}
In this work, we are interested in the dynamics of cell proliferation and differentiation in a heterogeneous cell population in the presence of different drugs which interact with each other. We represent cell counts in a vector $\mathbf{x} \in \mathbb{R}^n$ and the effective action of each of the drugs in a vector $\mathbf{u}\in\mathbb{R}^m$. At this point, we emphasize that the quantities $\mathbf{u}$ represent the pharmacodynamics of the different drugs, rather than their absolute concentration in the cell microenvironment. The advantage of this formulation is that we bypass the complicated functional forms expressing the pharmacokinetics, which are often highly problem-specific. For this reason, we assume that $0 \leq \mathbf{u}_k\leq 1$ for all $1 \leq k \leq m$. We describe the direct effect of a single drug on the differentiation and cell proliferation of cell type $j$ as a linear combination of all possible monomials in the form $\mathbf{u}_k\mathbf{x}_i$ for $1 \leq k \leq m$ and $1 \leq i \leq n$. We describe the effect of interactions between drugs using a minimal model whereby the governing equation for the $j$-th cell type contains a linear combination of all possible polynomials in the form $\mathbf{x}_i\mathbf{u}_k\mathbf{u}_{\ell}$. Here, we exclude terms containing $\mathbf{u}_k^2$ for every $1 \leq k \leq m$, as this is not a drug synergy. Denoting the terms containing products of$\mathbf{u}$ and $\mathbf{x}$ that are linear in $\mathbf{u}$ as $L(\mathbf{u}, \mathbf{x})$, and the terms that are nonlinear in $\mathbf{u}$ as $N(\mathbf{u}, \mathbf{x})$, we obtain that the dynamics of the vector $\mathbf{x}$ are described by an ordinary differential equation in the form
\begin{equation}
    \label{eq:introEquation}
    \dot{\mathbf{x}} = A\mathbf{x} + B\mathbf{u} + L(\mathbf{u}, \mathbf{x}) + N(\mathbf{u}, \mathbf{x}).
\end{equation}
Note that care must be taken with the sign of all polynomial terms to ensure that no population can become negative and the problem is well-defined. This equation will be coupled to a cost functional to yield an optimal control problem. 

\subsection{Functional form for the ODE model}
\label{section:FunctionalForm}
Here we explain how the terms containing products of the entries of $\mathbf{u}$ and $\mathbf{x}$ in Equation~\eqref{eq:introEquation} can be described with elementary matrix operations. For the terms that are linear in $\mathbf{u}$, we note that the model must account for monomials in the form $\mathbf{u}_k\mathbf{x}_i$ for $1 \leq k \leq m$ and $1 \leq i \leq n$ in the ODE for each of the $\mathbf{x}_j$ for $1 \leq j \leq n$. Hence, we introduce matrices $C_i$ for $1 \leq i \leq n$ such that
\begin{equation}
    \label{eq:LEquation}
    L(\mathbf{u}, \mathbf{x}) = \sum_{i=1}^n C_i \odot(\mathbf{1}_n \mathbf{x}^T\mathcal{E}_i)\mathbf{u},
\end{equation}
represents all the terms linear in $\mathbf{u}$ in the right hand side of Equation~\eqref{eq:introEquation}. Here we introduced the notation
\begin{equation*}
    \mathcal{E}_i = \mathbf{e}_i^n\mathbf{1}_m^T,
\end{equation*}
with $\mathbf{e}_i^n$ being the $i$-th standard basis vector in $\mathbb{R}^n$ and $\mathbf{1}_m \in \mathbb{R}^m$ a column vector containing only ones, \textit{i.e.} $\mathbf{1}_n\mathbf{x}^T\mathcal{E}_i$ is the $m\times n$ matrix containing only the $i$-th entry in $\mathbf{x}$. In other words, the entry $(C_i)_{k\ell}$ of the matrix $C_i$ contains the coefficient of the monomial $\mathbf{u}_\ell\mathbf{x}_i$ in the ODE for $\mathbf{x}_k$. For the terms containing interaction terms, \textit{i.e.}, products of the entries of $\mathbf{u}$, we again seek a functional form ensuring that the ODE for $\mathbf{x}_j$ for every $0 \leq j \leq n$ contains terms in the form $\mathbf{x}_i\mathbf{u}_k\mathbf{u}_{\ell}$. To this end, we introduce lower triangular matrices $D^{ij}$ and define
\begin{equation*}
    N(\mathbf{u}, \mathbf{x}) = \sum_{i=1}^n \sum_{j=1}^n (\mathbf{e}_j^n \mathbf{x}^T\mathcal{E}_i)(D^{ij} \odot \mathbf{u}\mathbf{1}_m^T)\mathbf{u},
\end{equation*}
where $\mathcal{E}_i$ is as above, and analogously to $L(\mathbf{u},\mathbf{x})$ in Equation~\eqref{eq:LEquation}, the entry $(D^{ij})_{k\ell}$ of the matrix $D^{ij}$ represents the coefficient of the term $\mathbf{x}_i\mathbf{u}_k\mathbf{u}_{\ell}$ in the ODE for $\mathbf{x}_j$. We introduce the convention that
\begin{equation*}
    (D^{ij})_{kk} = 0, \quad \text{for all} \quad 1\leq i,j \leq n,\quad 1 \leq k \leq m,
\end{equation*}
that is, we exclude quadratic terms in the action of any of the drugs in any of the governing equations, since we assume that the pharmacokinetics of any one drug are fully captured by the terms linear in $\mathbf{u}$. Finally, we note that $D^{ij}$ is lower triangular to represent symmetry in pairwise interactions. Put together, we can rewrite Equation~\eqref{eq:introEquation} as
\begin{equation}
    \label{eq:BasicEquation}
    \dot{\mathbf{x}} = A\mathbf{x} + B\mathbf{u} + \sum_{i=1}^n C_i \odot (\mathbf{1}_n \mathbf{x}^T \mathcal{E}_i)\mathbf{u} + \sum_{i=1}^n \sum_{j=1}^n (\mathbf{e}_j^n \mathbf{x}^T\mathcal{E}_i)(D^{ij} \odot \mathbf{u}\mathbf{1}_m^T)\mathbf{u},
\end{equation}
We finish the exposition of the functional form of our model by comparing it to another well-studied problem in optimal control, namely that of the linear quadratic regulator (LQR) \cite{Safaei12}. The equation for a LQR, which considers the effect of the control, $\mathbf{u}$, to be additive in the governing ODE of the state vector, $\mathbf{x}$, is given by
\begin{equation}
    \label{eq:LQR}
    \dot{\mathbf{x}} = A\mathbf{x} + B\mathbf{u}.
\end{equation}
It can be readily seen that our framework provides additional terms that include various nonlinear ways in which the control interacts with the governing equations. Given that an important appeal of the LQR is the ease with which the optimal control can be expressed as feedback control \cite{lenhart07optimalcontrol}, we will focus on demonstrating that the model extension considered in this problem lends itself to similar amenable properties for computation.

\section{Optimal control of an ODE with multiplicative controls}
\label{section:optimalControl}
Having formulated the governing equations for the model, we now turn to formulating an optimal control problem related to treatment response. In this continuous control problem, we consider a problem given in the form of Equation~\eqref{eq:BasicEquation}, and seek controls such that the corresponding solution of the model is an extremal of the functional
\begin{equation}
    J(u) = \frac{1}{2}\left[\mathbf{x}^T(T)M\mathbf{x}(T) + \int_0^T\left\{\mathbf{x}^T(t)Q(t)\mathbf{x}(t) + \mathbf{u}^T(t)R(t)\mathbf{u}(t)\right\}\dd t\right].  \label{eq:costFunction}
\end{equation}
As is standard in optimal control theory, $M, Q$ and $R$ are symmetric matrices, where $M, Q$ are positive semi-definite, and $R$ is positive definite for all $t \in [0,T]$ \cite{lenhart07optimalcontrol}. The interpretation of the different terms of the cost function in Equation~\eqref{eq:costFunction} is as follows. The matrix $M$ associates a cost to the terminal entries of the state. In mathematical oncology models, for example, this could represent the final tumor population. The matrix $Q$ describes the cost associated with the state across the whole interval $[0,T]$. The matrix $R$ describes costs associated with the drug $\mathbf{u}$. In models describing treatment, this term is often taken to represent treatment toxicity or cost \cite{lenhart07optimalcontrol}.

Having introduced the governing equation in Equation~\eqref{eq:BasicEquation} together with its cost functional in Equation~\ref{eq:costFunction}, it is possible to follow standard arguments in optimal control theory to formulate the conditions for the control to be optimal. We first formulate a Hamiltonian, $H$, associated with the problem,
\begin{equation}
    H = \frac{1}{2}\left[\mathbf{x}^T(t)Q(T)\mathbf{x}(t) + \mathbf{u}^T(t)R(t)\mathbf{u}(t)\right] + \boldsymbol{\lambda}\cdot\dot{\mathbf{x}}.
\end{equation}
Here, $\boldsymbol{\lambda}$ is the vector of adjoints, or co-states, and it satisfies the system of equations
\begin{equation}
    \label{eq:Adjoint}
    \dot{\boldsymbol{\lambda}}_k = - \frac{\partial H}{\partial \mathbf{x}_k},
\end{equation}
for all $k \in \{1,\dots,n\}$. To close the system of equations, we impose the optimality condition ensuring that the control is in fact a minimizer of the cost functional. This optimality condition is given by 
\begin{equation}
    \label{eq:Optimality}
    \frac{\partial H}{\partial \mathbf{u}_k} = 0,
\end{equation}
for all $k \in \{1,\dots,n\}$ \cite{lenhart07optimalcontrol}. We remark here that in most well-posed problems, the optimality condition provides a means to algebraically express $\mathbf{u}$ in terms of $\boldsymbol{\lambda}$ and $\mathbf{x}$. In some cases, which we will not consider in this paper, however, $\partial H /\partial \mathbf{u}_k$ does not depend on $\mathbf{u}$. In this case, several techniques exist to address the problem \cite{lenhart07optimalcontrol}.

\subsection{Adjoint equations}
The adjoint equation in Equation~\eqref{eq:Adjoint} can be expressed in terms of the model coefficients defined in Equation~\eqref{eq:BasicEquation}. The derivation in Appendix~\ref{appendix:Adjoint} shows that the adjoint equation in this problem is given by
\begin{equation}
    \dot{\boldsymbol{\lambda}} = -Q\mathbf{x} - A^T\boldsymbol{\lambda} - \sum_{i=1}^n \mathbf{e}_i^n (C_i\mathbf{u})^T \boldsymbol{\lambda} - \sum_{i=1}^n \mathbf{e}_i \left(\sum_{j=1}^n \mathbf{e}_j^n\mathbf{1}_m^T (D^{ij}\odot \mathbf{u}\mathbf{1}^T_m)\mathbf{u}\right)^T\boldsymbol{\lambda}.
\end{equation}
\subsection{Optimality conditions}
The optimality condition in Equation~\eqref{eq:Optimality} (see Appendix~\ref{appendix:Optimality}) gives rise to
\begin{equation}
    \label{eq:OptimalClosedForm}
    \mathbf{u}^{\star} = -\left(R+\sum_{i=1}^n\sum_{j=1}^n (\mathbf{e}_i^n)^T \mathbf{x}\boldsymbol{\lambda}^T \mathbf{e}_j^n D^{ij}\right)^{-1}\left(B+\sum_{i=1}^n C_i \odot (\mathbf{1}_n \mathbf{x}^T \mathcal{E}_i)\right)^T\boldsymbol{\lambda},
\end{equation}
whenever the inverse in Equation~\eqref{eq:OptimalClosedForm} exists. 
If we compare this solution to the optimal control in the case of the LQR, we find
\begin{equation} 
    \label{eq:LQRcontrol}
     R\mathbf{u} + B^T\boldsymbol{\lambda} = \mathbf{0}.
\end{equation}
Hence, we observe that both the multiplicative control and the drug-drug interaction terms contribute additional terms compared to the optimal control of the LQR. Moreover, these corrections depend directly on both the state and the co-state variables, $\mathbf{x}$ and $\boldsymbol{\lambda}$, respectively. Note that the matrix in the first term of the optimal control as defined in Equation~\eqref{eq:OptimalClosedForm} cannot always be inverted, as opposed to that of the control in the LQR in Equation~\eqref{eq:LQRcontrol}, for which an inverse always exists. In such cases, one needs to resort to Pontryagin's optimality criterion, or the solution of singular optimal control problems, which can be avoided in specific examples \cite{lenhart07optimalcontrol}. Having established a framework to perform optimal control in a system consisting of a heterogeneous cell population with multiple interacting drugs, we focus on showcasing the utility and applicability of the method on three illustrative problems.

\section{Example 1: Controlling growth in a two-population model}
\label{section:example1}
\begin{figure}
    \centering
    \includegraphics[width=0.5\linewidth]{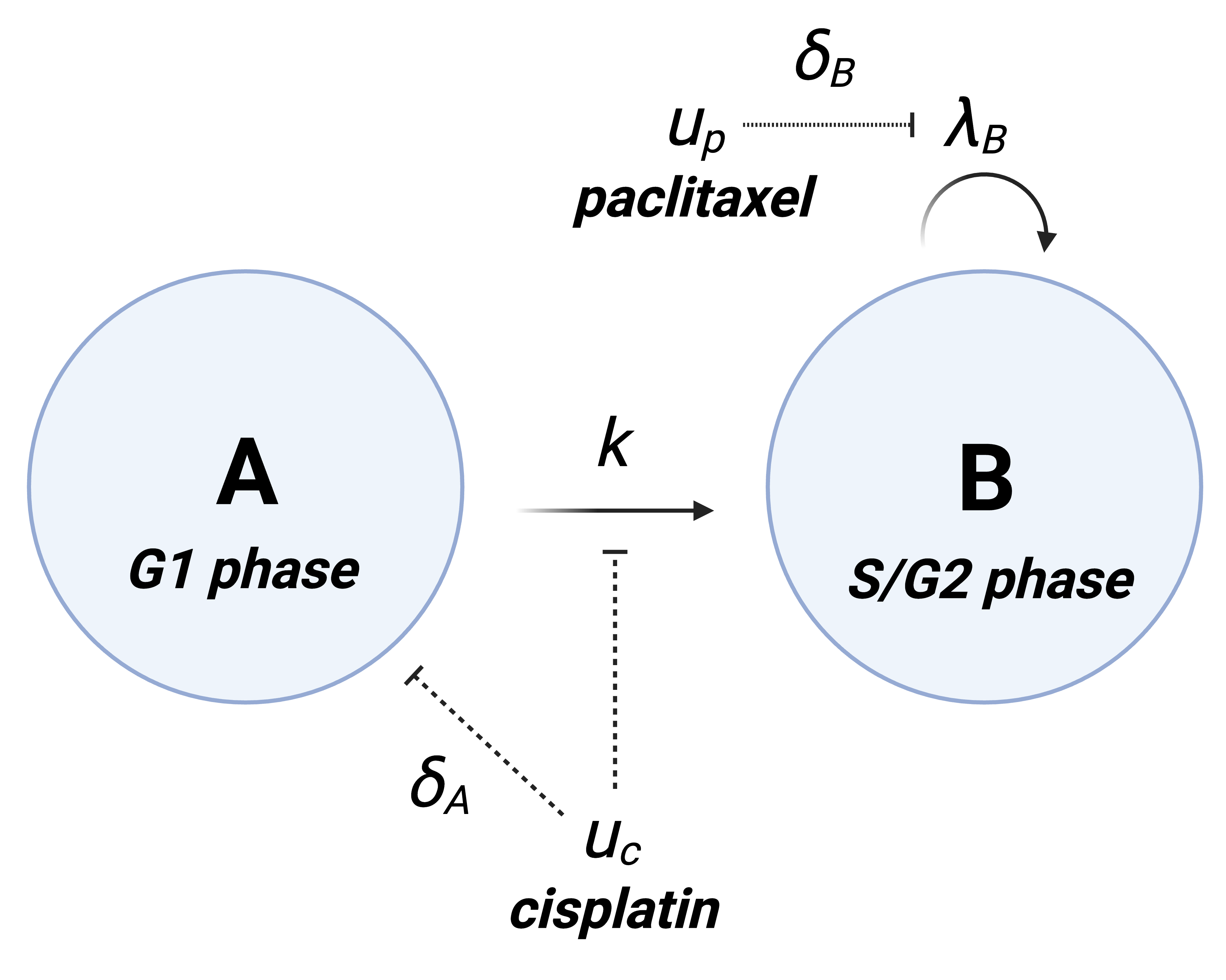}
    \caption{ Schematic of the two-population ODE model. Cervical cancer cells are divided into two populations, A and B, according to their phase of the cell cycle. G1 phase (non-proliferative) cells are denoted A, whilst S/G2 phase (proliferative) cells are denoted B. Cells enter the cell cycle at rate $k$. Cisplatin (concentration denoted by $u_c$), reduces the rate of transfer of population B to population A and kills cells in the G1 phase at rate $\delta_A$. Paclitaxel (concentration denoted by $u_p$), reduces the rate of proliferation in population B and leads to cell death at rate $\delta_B$. }
    \label{fig:twoPopSchem}
\end{figure}

In cancer treatment, drug efficacy is known to vary with the phase of the cell cycle that a cell occupies \cite{shah2001cell, lieftink2021takes}. Drug efficacy dependence on the cell-cycle offers a potential means of maximizing the efficacy of therapies whilst minimizing drug dosing, and hence the adverse effects on healthy tissue \cite{lieftink2021takes}. In this example, we explore the possible synergies of two drugs in a simplified system consisting of cervical cancer cells. We distinguish between cells in the G1 phase (denoted by A in Figure~\ref{fig:twoPopSchem}) and those in the S/G2 phase (denoted by B in Figure~\ref{fig:twoPopSchem}). The cells are treated with a combination of paclitaxel (with pharmacodynamics denoted by $u_p$) and cisplatin (with pharmacodynamics denoted by $u_c$). Paclitaxel disrupts the normal microtubule dynamics required for cell division, and hence, arrests cell division in the S/G2 phase \cite{markman2002paclitaxel}. Owing to the arrest in the G2/S phase, cell death occurs in a process known as mitotic catastrophe \cite{markman2002paclitaxel}. Cisplatin, on the other hand, prevents DNA replication, leading to synchronization of the cell cycle in the G1 phase \cite{romani2022cisplatin}. Cells that fail to properly replicate their DNA undergo apoptosis as a result \cite{romani2022cisplatin}. The combination therapy of cisplatin and paclitaxel has been shown to be more favourable in the treatment of stage III and stage IV cervical cancer than single-drug cisplatin or paclitaxel treatments \cite{muggia2000phase}. However, both drugs are associated with significant side effects: fever and alopecia commonly occur with paclitaxel, and neurotoxicity and anemia commonly occur with cisplatin. As such, it is desirable to minimize the dose of both drugs given to a patient whilst also minimizing the population of cancer cells. 

\subsection{ODE model}
Assuming exponential growth of each population, and writing the number of cells in population A by $n_A$ and the number of cells in population B as $n_B$, the dynamics between drugs and cells represented in Figure~\ref{fig:twoPopSchem} can be expressed with the following system of coupled ODEs
\begin{align*}
    \frac{\dd n_A}{\dd t} &= 2\lambda_B (1-u_p) n_B - kn_A(1-u_c) - \delta_A u_c n_A,\\
    \frac{\dd n_B}{\dd t} &= -\lambda_B(1-u_p)n_B - \delta_B u_p n_B + kn_A(1-u_c).
\end{align*}
Here, $\lambda_B$ is the rate of division for cells in the G2/S phase, $k$ is the rate at which cells transition from the G1 phase to the G2 phase, $\delta_A$ is the rate of cell death induced by cisplatin for cells in the G1 phase, and $\delta_B$ is the rate of cell death induced by paclitaxel for cells in the G2/S phase. The dimensionless quantities $u_\bullet \in [0, 1]$ represent the concentration of the different drugs in the system. Since cervical cancer cells spend roughly the same time in the G1 and the G2/S phase, we set $\lambda_B = k$. Non-dimensionalising the system by taking $\tau = kt$, $N_A = kn_A$, and $N_B= \lambda_B  n_B$, one obtains the dynamics
\begin{align}
    \frac{\dd N_A}{\dd \tau} &= 2N_B(1-u_p) - N_A (1-u_c) - \alpha u_c N_A \label{coupled1}, \\
    \frac{\dd N_B}{\dd \tau} &= -N_B(1-u_p) + N_A(1-u_c) - \beta u_p N_B, \label{coupled2}
\end{align}
with $\alpha = \delta_A/k$ and $\beta = \delta_B/k$. 

Having expressed the model in non-dimensionalized form, the relative importance of terms in Equations \eqref{coupled1} and \eqref{coupled2} is readily determined by the values of the dimensionless parameter groupings, $\alpha$ and $\beta$. Furthermore, the dynamics of the system are now described on the timescale for the rate of differentiation of cells in G1 (type A) into cells in G2/S phase (type B). 

\subsection{Optimal control formulation}
To express the cost function associated with the treatment, we set the penalty for the cell counts equal to the identity, \textit{i.e.}, $Q = \mathbb{I}$, and the penalty for the drug toxicity equal to $R = 10^{-1} \mathbb{I}$. Equations~\eqref{coupled1} and \eqref{coupled2} can be represented into the matrix form of Equation~\eqref{eq:BasicEquation} by setting
\begin{equation*}
    A = \begin{pmatrix}
        -1 & 2 \\ 1 & -1
    \end{pmatrix}, \quad C_1 = \begin{pmatrix}
        1- \alpha & 0 \\ -1 & 0
    \end{pmatrix}, 
    \quad C_2 = \begin{pmatrix}
        0 & -2 \\ 0 & 1 - \beta
    \end{pmatrix}.
\end{equation*}
Since there are no drug-drug interactions in this example, it follows that $D^{ij} = 0$, for all $i ,j$. In the general case presented in the previous section, the optimal control defined in Equation~\eqref{eq:OptimalClosedForm} is coupled with the adjoint and state equations. In this example, however, the fact that there are no drug-drug interactions, \textit{i.e.}, $D^{ij} = 0$, means that the optimal control as defined in Equation~\eqref{eq:OptimalClosedForm} always exists, with
\begin{equation}
    \label{eq:ReducedOptimalControl}
    \mathbf{u}^{\star} = -R^{-1}\left(\sum_{i=1}^n C_i \odot (\mathbf{1}_n \mathbf{x}^T \mathcal{E}_i)\right)^T\boldsymbol{\lambda},
\end{equation}
subject to the condition that $0 \leq \mathbf{u}_k(t) \leq 1$, for all $k$ and all $t \geq 0$. The functional form of the control in Equation~\eqref{eq:ReducedOptimalControl} resembles that encountered in the LQR problem. One might think that in this problem, therefore, it is possible to cast the optimal solution $\mathbf{u}^*$ as a feedback control. Such a control would be expressed in the state vector alone by eliminating the co-state variable from the control equation through the use of a Ricatti equation \cite{lenhart07optimalcontrol}. This, however, is not possible due to the coupling of the control, $\mathbf{u}$, and its bounds, with the governing equation for the co-state, $\boldsymbol{\lambda}$. For the benefit of the reader we demonstrate this in Appendix~\ref{appendix:RicattiEquation}.

\subsection{Numerical implementation}
\label{section:numericalImplementation}
We numerically solve the boundary value problem arising from the initial condition for the cell populations and the transversality condition, given by $\lambda_{\bullet}(T) = 0$, where $\bullet=B$ in the above example. To this end, in each simulation we employ the bvp4c scheme of Kierzenka \textit{et al.} implemented in the \verb|SciPy| package in Python~\cite{virtanen20scipy}. This scheme implements a fourth-order collocation algorithm with control of the residuals and uses a damped Newton method with an affine-invariant criterion function. For the implementation, we set a maximum relative tolerance of $10^{-3}$, an absolute boundary value tolerance of $10^{-8}$ and specify a maximum of $5 \times 10^5$ nodes for the collocation algorithm. Example code has been made available for all computations at our GitHub repository: \url{https://github.com/SWSJChCh/multiplicativeControl}.

\subsection{Results}
Having two parameters in the model given by Equations~\eqref{coupled1} and \eqref{coupled2}, we can explore the effect of varying system parameters on the resulting optimal controls. For all simulations, we start with an equal population of cells in G1 and S/G2: $N_A = N_B = 1$, and simulate a standard one week course of chemotherapy, which in non-dimensional units is roughly equal to $T = 7$ (since time has been re-scaled by the typical cell cycle length, which is roughly 24 hours). The first question we investigate is how effective the optimal control is in reducing the cell count at the end of the treatment interval, which is a clinically relevant question when it comes to improving treatment outcomes. Naively, one might expect the optimal control to always outperform a solution with constant pharmacodynamics with the same area under the curve. \textit{A priori}, there is no reason to assume that this is guaranteed: the optimal control is defined to minimize the cost functional in Equation~\eqref{eq:costFunction}, which considers the dynamical trajectory of both the cell counts and pharmacodynamics. The following discussion serves to underscore the wider point that optimal control can be useful in finding pharmacodynamics that outperform naive constant dosing treatment, as we will show. This demonstrates that one could in fact use optimal control to find solutions that are not just mathematically optimal according to Equation~\eqref{eq:costFunction}, but are also practically useful. 

\begin{figure}[htb]
    \centering
    \includegraphics[width=\linewidth]{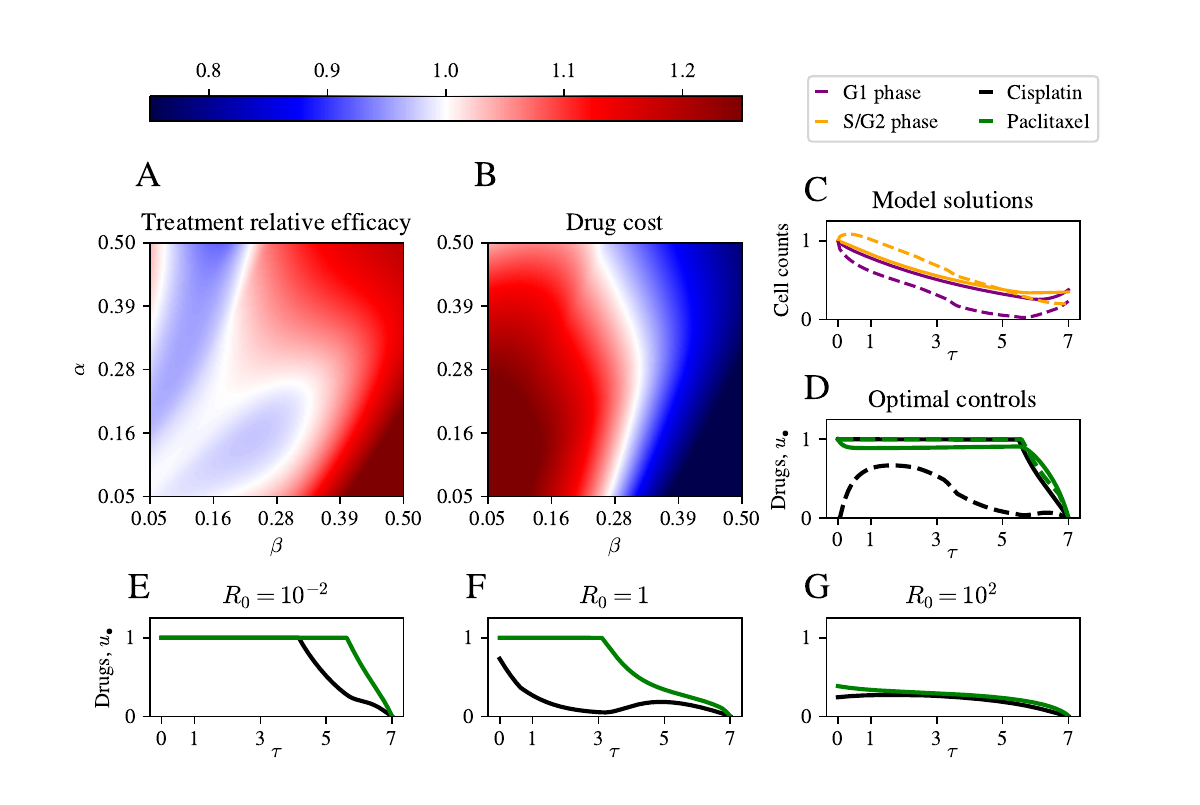}
    \caption{Efficacy of optimal control treatment in two-population model. A: Heat map of relative treatment efficacy, $\eta$ (as defined in Equation~\eqref{eq:eta}) for different system parameters, $\alpha$, and $\beta$. B: Total integrated cost of the drug administered. C: model solutions using optimal control at $\alpha = 0.05$, $\beta = 0.5$ (dashed line) and $\alpha = 0.5$, $\beta = 0.05$ (solid line). D: Optimal controls in the same parameter regimes as panel C. E-G: Optimal controls for $\alpha = \beta = 0.5$ and varying $R_0$ (the entry along the leading diagonal of the cost matrix $R$).}
    \label{fig:ex1EqualToxicity}
\end{figure}

To make a systematic and fair comparison between an optimal control treatment and that with constant pharmacodynamics, we simulate different model trajectories for ${\alpha, \beta \in [0.05, 0.5]}$ and compare each of the solutions arising from the optimal control to treatment with constant pharmacodynamics defined by 
\begin{equation}
    \bar{\mathbf{u}}_\bullet = \frac{1}{T}\int_0^T \mathbf{u}_\bullet(\tau) \dd \tau,
\end{equation}
for each individual drug, $\mathbf{u}_\bullet$. This constant administration regime yields a corresponding solution, $\bar{\mathbf{x}}$. We then measure efficacy of the combination treatment by using the ratio
\begin{equation}
    \label{eq:eta}
    \eta = \frac{\sum_{i=1}^n \bar{\mathbf{x}_{i}}(T)}{\sum_{i=1}^n \mathbf{x_{i}}(T)}.
\end{equation}
The heat map in Figure~\ref{fig:ex1EqualToxicity} (panel A) shows that there is a complicated and non-linear relationship between system parameters and the efficacy of the drug. For higher values of $\beta$, \textit{i.e.}, for higher paclitaxel drug kill, the optimal solution outperforms the constant control significantly. On the other hand, when cisplatin drug kill is high and that of paclitaxel is low, the constant drug regime performs better. Therefore, the optimal control framework leads to practical improvements in treatment outcome by taking into account temporal variations in pharmacodynamics in the high paclitaxel drug kill regime, whereas the benefit of doing so in the high cisplatin drug kill regime is not present. To avoid confusion, at this point we stress that the optimal control as defined by the cost functional $J$ in Equation~\eqref{eq:costFunction} produces a trade-off between drug toxicity and cell counts during the experiment. For this reason, it cannot be immediately expected that the optimal control will always outperform a constant drug dose when comparing only final cell counts between different administration regimes. 

To tease apart the possible mechanisms that could explain why in some parameter regimes the constant treatment is more effective than the optimal control, or \textit{vice versa}, we turn to the question of how much drug is administered when the system parameters are varied. The heat map in Figure~\ref{fig:ex1EqualToxicity} (panel B) shows the integrated drug cost, \textit{i.e.} $\int_0^T \mathbf{u}^T R \mathbf{u}\dd\tau$, for different parameter values. This figure shows that for higher values of paclitaxel drug kill, $\beta$, the drug cost decreases. The cost also decreases for lower values of cisplatin drug kill, $\alpha$. Put together with the findings in panel A of Figure~\ref{fig:ex1EqualToxicity} we find that, generally, for small circulating amounts of the drug, varying the treatment with time according to the optimal control prediction is more effective that simply applying a constant, averaged treatment. Here, this might suggest that, when only a small amount of drug is administered, the temporal distribution of the drug significantly impacts treatment response. Finally, we remark that the optimal control terminal condition, $\lambda_\bullet(T) = 0$, was used to ensure optimality of the control, but in our model translates to the drugs being withheld at the end of the treatment window. For this reason, we can expect the corresponding model solutions to show a modest increase in the drug effect, $\boldsymbol{u}$, in the last part of the treatment window, whereas this is not the case for the constant control, which maintains non-zero treatment across the entire window.

Finally, in the preceding analysis we have noted that the optimal control formulation using the cost function $J$ in Equation~\eqref{eq:costFunction} makes a trade-off between reducing the total cell count and reducing the toxicity associated with drug treatment. For this reason, we now explore how the optimised treatment protocols are influenced when model parameters are kept the same, but the matrix $R$ controlling drug toxicity is varied. To do this, we take $\alpha = \beta = 0.5$ and let the drug toxicity vary by considering multiples of the identity matrix, \textit{i.e.} $R = R_0\cdot\mathbf{I}$, where $R_0$ is a constant. We see in Figure~\ref{fig:ex1EqualToxicity} (panels E-G) that increasing $R_0$ from $R_0 = 10^{-2}$ to $R_0 = 10^2$ has a dramatic impact on the dynamics of the optimal control. For low drug toxicity, $R_0 = 10^{-2}$, the cost functional $J$ allows for possibly large amounts of drug administration, leading to a near-constant drug profile (panel E). When $R_0$ is increased to $R_0 = 1$, the cost formulation leads to a trade-off where one of the two drugs is favoured. Finally, when $R_0$ is further increased to $R_0 = 10^2$, the amount of drug is drastically reduced for both drugs. We conclude that calibrating the drug penalty is therefore crucial for possible biomedical applications of optimal dosing, although it remains an open question as to how this should be done, and how the parameter $R_0$ can be measured experimentally.

\begin{figure}[htb]
    \centering
    \includegraphics[width=\linewidth]{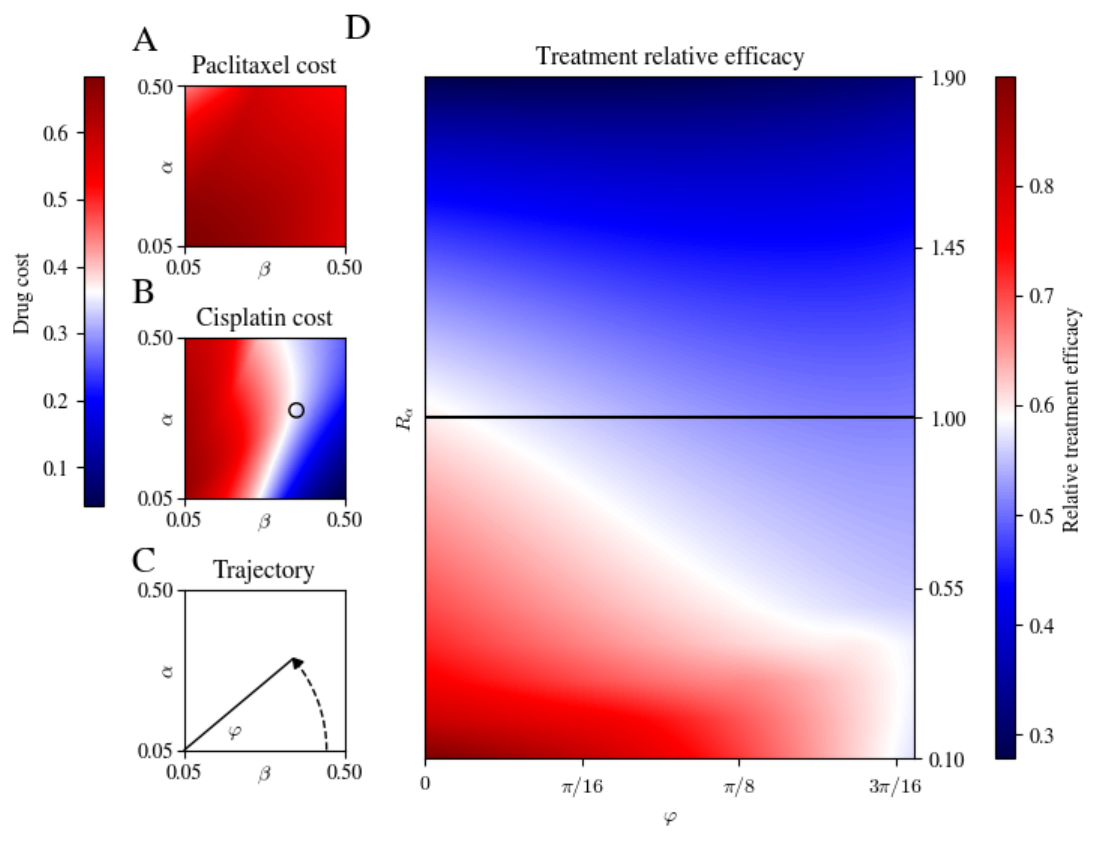}
    \caption{Efficacy of treatment upon varying drug toxicity. A: Heat map showing the cost incurred by paclitaxel administration with constant cost matrix as in Figure~\ref{fig:ex1EqualToxicity}. B: Heat map showing cost incurred by cisplatin administration. Black circle indicates the endpoint of the trajectory in parameter space visualised in D. C: Parameterised trajectory through parameter space along curve with significant cisplatin cost variation. D: Heat map showing treatment efficacy along the path in panel C as relative cisplatin toxicity is varied.}
    \label{fig:ex1VaryingToxicity}
\end{figure}

In the preceding analysis, we have used a cost functional that penalizes both drugs, cisplatin and paclitaxel, equally. However, in many clinical settings it might be more desirable to avoid the use of one drug rather than another. This might be due to the fact that one of them is more toxic, but also it is possible that specific patients tolerate one drug better than the other. To investigate the role of varying the relative toxicity of the drugs, we introduce a model extension to vary the relative penalty of one drug versus another. We do this by letting the cost matrix $R$ be given by 
\begin{equation*}
    R = \begin{pmatrix}
        R_\alpha &0 \\
        0 & R_\beta
    \end{pmatrix}, \quad R_\alpha + R_\beta = R_0, \quad R_\alpha, R_\beta > 0,
\end{equation*}
where $R_0$ is given by $R_0 = 2\times 10^{-1}$, \textit{i.e.}, as in the standard case in the preceding analysis. We then investigate which regions of parameter space incur the highest cost due to either cisplatin or paclitaxel administration (Figure~\ref{fig:ex1VaryingToxicity}, panels A and B). 

In the heat maps in Figure~\ref{fig:ex1VaryingToxicity} (panels A and B) we note that there are different regions of parameter space in which the cost arising from administration of either of the two drugs varies. To investigate how this balance changes when we penalize administration of one drug more than the other, we set a path through parameter space which shows high variability in cisplatin cost, which for ease we parameterize as
\begin{equation}
    \alpha = \gamma_0 + \gamma \sin(\varphi), \quad \beta = \gamma_0 + \gamma \cos(\varphi), \quad \varphi \in \left[0,\frac{2\pi}{9}\right],
\end{equation}
with $\gamma_0 = 0.05$ and $\gamma = 0.4$, where the maximum value of $\varphi$ corresponds to the approximate bifurcation point in Figure~\ref{fig:ex1VaryingToxicity}B. By varying, $R_\alpha$ and $\varphi$, we have a means to investigate the role of penalizing the different drugs as we move through regions of parameter space that favour one drug more than the other. The heat map in Figure~\ref{fig:ex1VaryingToxicity} (panel D) shows that treatment efficacy increases when toxicity for cisplatin is reduced compared to that of paclitaxel. This is more pronounced for larger values of $\varphi$, \textit{i.e.}, closer to the $\alpha$ axis in the schematic of Figure~\ref{fig:ex1VaryingToxicity} (panel C). That is, when the tolerance for cisplatin is increased in a parameter region that flavors the administration of this drug, it can be expected that the optimal drug dosing regime will allow more for it to be administered, with a favourable effect on treatment outcome. 

Put together, we have shown that optimal control can be used to formulate a time-dependent dosing regime that can outperform naive treatment with constant pharmacodynamics. This is not intuitive \textit{a priori}, but is useful because it allowed to obtain better treatment outcomes in the simulations of Figure~\ref{fig:ex1EqualToxicity} when only small amounts of drug were administered. Therefore, the wider utility of optimal control in drug treatment may include finding bespoke temporally varying dosing regimes of different drugs when only very little drug can be administered, for example in contexts of high drug toxicity, where one might otherwise consider a constant administration of very little drug. In addition, a careful calibration of the relative toxicity of the drugs, can have a dramatic impact on the relative amounts of drug administered, as well as the overall efficacy of the drug treatment. Together, we have presented a mathematical case study in how to explore the utility and applicability of optimal control to a treatment problem. 

\section{Example 2: Modeling neuroblastoma differentiation and regression}
\label{section:example2}
In this example, we study a control problem arising from the treatment of neuroblastoma -- a pediatric cancer in which cellular heterogeneity plays a large role in disease outcomes \cite{Cohn09, Zeineldin22awry}, and for which several different drug treatments exist \cite{Zeineldin22awry, Zhou23Resistance}.  We demonstrate in this section that, while a realistic ODE model is highly complicated and contains many different terms, optimal control theory can be used to tease apart the many different responses of the cell population. In the context of using optimal control for treatment design, we use this as a case study to address the general point of exploring possible treatment strategies when the number of drugs is very large, and the possible temporal dynamics of drug administration are mathematically intractable. 

\subsection{Neuroblastoma: cellular heterogeneity and the extracellular environment}
\label{section:neuroblastomaBackground}
Neuroblastoma is a extra-cranial solid cancer which is thought to originate from a neural-crest derived progenitor in children \cite{Zeineldin22awry, Korber23, Jansky21Origin, Gomez22}. Neuroblastoma tumors originate early in fetal and infant development \cite{Korber23} and are thought to arise due to a differentiation arrest of the early sympathoadrenal lineage \cite{Zeineldin22awry}. The result of this differentiation arrest is the presence of tumors along the sympathetic nervous system containing cells in an immature cell differentiation state \cite{Zeineldin22awry, Gomez22}. This heterogeneity has important consequences for drug treatment \cite{Zeineldin22awry, Gomez22}.  In the rest of this section, we will provide a thorough overview of cell types, drugs, and their interactions, to thoroughly motivate an ODE model for optimal control. 

A common framework of the cell types most frequently found within neuroblastoma tumours classifies the cells in a neuroblastoma tumor as sympathoblasts (I-type), adrenergic (N-type), or mesenchymal (S-type) \cite{Zeineldin22awry}. I-type cells are proliferative cells that originate from the neural crest. Due to their multi-potency, they can produce both N-type and S-type cells \cite{Zeineldin22awry, Jansky21Origin}. Immature I-type cells are thought to be a potential reservoir for malignant cells to expand and form neuroblastoma tumours \cite{Jansky21Origin, Korber23}. N-type cells are further differentiated into cells that appear neuroblastic, and can differentiate into neuronal tissue in the presence of retinoic acid (RA) \cite{Zeineldin22awry}. Terminally differentiated neuronal cells are minimally proliferative \cite{Zeineldin22awry}. S-type cells are substrate-adherent, express vimentin, and produce collagen as well as fibronectin \cite{Zeineldin22awry, Jansky21Origin}. S-type and N-type neuroblastoma cells can inter-convert between cell types \cite{Zeineldin22awry, Groningen17}. 

While the exact role of each cell type in the development of neuroblastoma is still largely unclear \cite{Zeineldin22awry, Gomez22, Maris10, Schmelz21}, there are some general clinical themes. All three cell types have been found in neuroblastoma tumours, but their relative abundance is highly variable between patients \cite{Gomez22}. Patients with a higher abundance of differentiated N-type cells have overall better prognoses \cite{Zeineldin22awry}. Tumor heterogeneity therefore has important consequences for drug treatments. Various drug treatments exist for neuroblastoma whereby drugs are used on their own, or in addition to surgical resection \cite{Maris10, Cohn09, Brodeur14Regression, Schmelz21}. A first treatment is the induction of neuronal differentiation as a therapeutic. Treatment with RA has been shown to lead to irreversible differentiation of I-type cells into N-type cells \textit{in vitro} and \textit{in vivo} \cite{Matthay99}. For this reason, the RA derivative 13-cis RA is now part of the standard of care for neuroblastoma \cite{Zeineldin22awry}. We note, however, that S-type cells are not sensitive to treatment with RA \cite{Gomez22}. A second treatment uses a cytotoxic chemotherapeutic agent and is given in several cycles before or after surgery \cite{Matthay16}. Being less proliferative, N-type cells are significantly less sensitive to treatment with a chemotherapeutic agent, and S-type cells are widely understood to be resistant to chemotherapy \cite{Zeineldin22awry, Gomez22, Boeva17}.

A feature of neuroblastoma is that cells in the sympathoadrenal lineage respond to environmental signals that drive differentiation and development, which is in large part mediated by the tropomyosin receptor kinase family (trk) \cite{Zeineldin22awry, Kasemeier18Logic, Kasemeier-Kulesa2015TrkB/BDNFSystem}. The first member of the trk family is trkA, which has an important role in cellular differentiation. It is believed that I-type cells that express trkA and are in an environment with the extracellular factor NGF differentiate more readily into N type cells \cite{Kasemeier18, Brodeur14Regression}. When N-type cells, in contrast, express trkA but are not in contact with NGF, they undergo apoptosis \cite{Brodeur14Regression}. The second member of the trk family is trkB, which is known to have an important role in disease progression \cite{Kasemeier-Kulesa2015TrkB/BDNFSystem}. In brief, N-type cells in the presence of trkB and the extracellular factor BDNF can de-differentiate into I-type or S-type cells \cite{Brodeur14Regression, Kasemeier18, Kasemeier-Kulesa2015TrkB/BDNFSystem, Kasemeier18Logic}. Currently, treatments exist that inhibit receptors of the trk family using pan-trk inhibitors (\textit{i.e.} they inhibit both trkA and trkB simultaneously) \cite{Brodeur14Regression,Gomez22}. An example of a pan-trk inhibitor used in neuroblastoma is entrectinib \cite{Iyer16}. The trk signaling pathways are complicated to model since they interact with other factors present -- in this case NGF and BDNF -- the concentration of which can vary according to the age of the patient and the anatomical location of the tumor \cite{Kasemeier-Kulesa2015TrkB/BDNFSystem, Kasemeier18Logic}. Therefore, the effect of any one drug on trk inhibition will also depend on other chemical species present, leading to nonlinear effects in Equation~\eqref{eq:BasicEquation}, which provides an important extension to previous work in our optimality condition.

\begin{figure}[htb]
    \centering
    \includegraphics[width=0.65\linewidth]{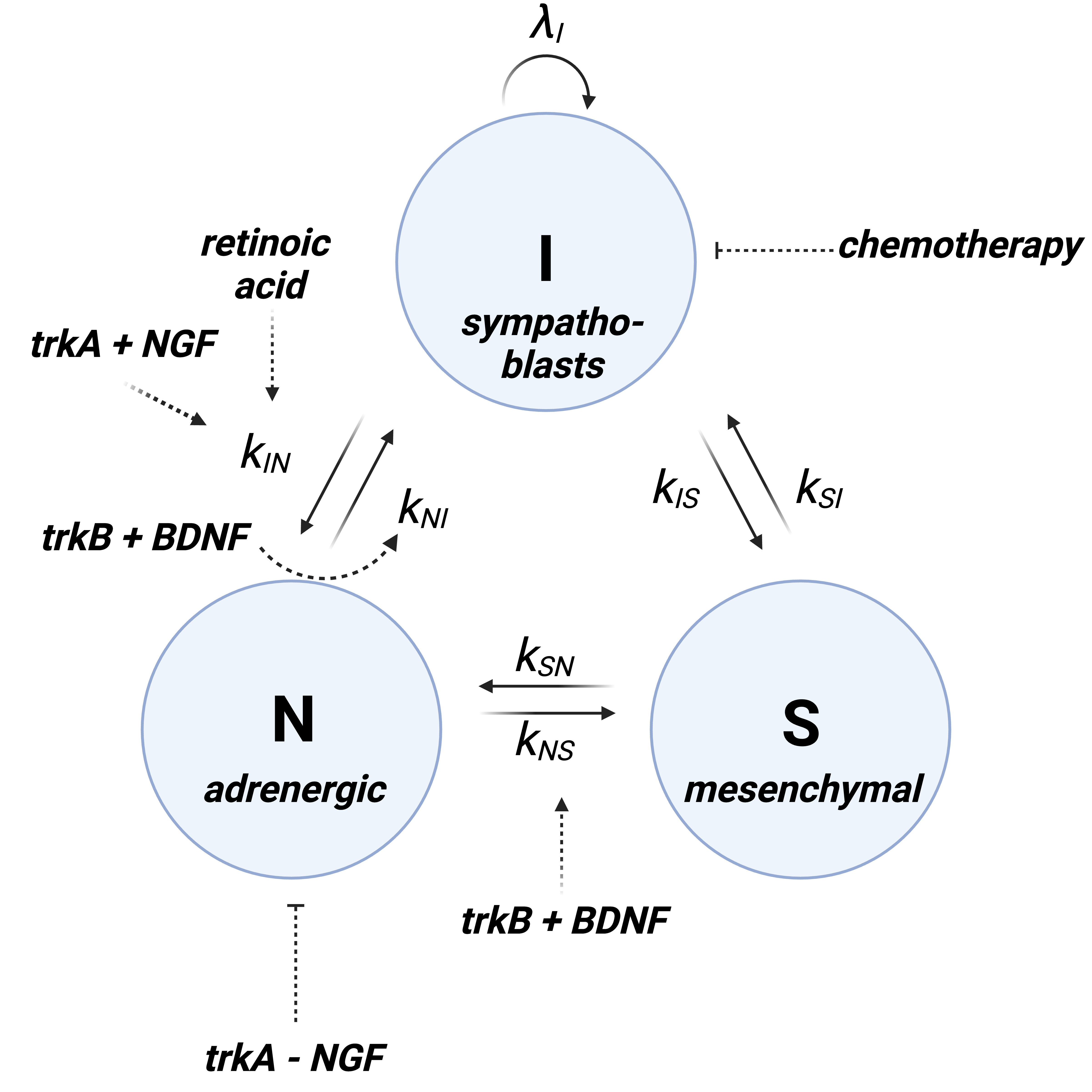}
    \caption{Schematic of the three-population neuroblastoma ODE model. Cells are divided into I-type sympathoblasts, N-type adrenergic cells, and S-type mesenchymal cells. The model includes spontaneous interconversion between all cell types, treatment with RA acting on the I-to-N differentiation pathway, chemotherapy affecting I-type cells, combined trkA-NGF signaling affecting I and N-type cells, and trkB-BDNF signaling affecting N and S-type cells.}
    \label{fig:NeuroblastomaSchematic}
\end{figure}

We remark that there are many other treatments for neuroblastoma that selectively target specific subpopulations, such as anti-GD2 immunotherapy, or treatment with metaiodobenzyl- guanidine (MIBG) to name two examples \cite{Zeineldin22awry}. These treatments can be added to the model if desired. For the ease of exposition, we have opted to limit the discussion to one drug per effect. We summarize this system containing three cell populations and drug interactions in Figure~\ref{fig:NeuroblastomaSchematic}. In the remainder of this example, we formulate a system of ODEs describing these multi-drug interactions in Section~\ref{section:neuroblastomaODE}, describe the corresponding optimal control problem in Section~\ref{section:neuroblastomaOptimalControl}, and provide insights into drug synergies in neuroblastoma drug treatment in Section~\ref{section:neuroblastomaInsights}.

\subsection{A multi-drug, multi-population model of cell proliferation, differentiation, and death in neuroblastoma}
\label{section:neuroblastomaODE}
In this section, we formulate a system of ODEs describing the multi-drug and multi-cell population dynamics described in Section~\ref{section:neuroblastomaBackground}. Denoting the cell numbers of the I-type, N-type and S-type cell populations as $\nI, \nN, \nS$, respectively, we introduce first the governing equation for the number of I-type cells, $\nI$:
\begin{equation} \label{eq:nIEquation}
\begin{split}
\frac{\dd \nI}{\dd t} = \overbrace{\lambda_{\text{I}}\nI}^{\text{proliferation}} &- \overbrace{(k_{\text{IN}}+k_{\text{IS}})\nI + k_{\text{NI}}\nN + k_{\text{SI}}\nS}^{\text{spontaneous interconversion}} -\overbrace{\delta_{\text{RA}}u_{\text{RA}}\nI}^{\text{RA-induced diff.}} - \overbrace{\delta_{\text{chemo}}u_{\text{chemo}}\nI}^{\text{chemo-induced apoptosis}} \\
    &- \underbrace{\delta_{\text{diff}}(1-u_{\text{trk}})u_{\text{NGF}}\nI }_{\text{trkA+NGF induced diff.}} + \underbrace{\delta_{\text{BDNF}}(1-u_{\text{trk}})\nN.}_{\text{trkB+BDNF induced de-diff.}}
\end{split}
\end{equation}
Here, $\lambda_{\bullet}$ represents the proliferation rate of $\bullet$-type cells, the constants ${k_{\bullet\dagger} \geq 0}$ represent the rates of spontaneous interconversion between cell type $\bullet$ and $\dagger$. The constants $\delta_{\text{RA}}$, $\delta_{\text{chemo}}$, $\delta_{\text{diff}}$, and $\delta_{\text{BDNF}}$ correspond to the effects of retinoic acid on differentiation, chemotherapy on cell death, trkA on cell differentiation, and BDNF on de-differentiation, respectively. For the governing equation for the number of N-type cells, $\nN$, we write
\begin{equation} \label{eq:nNEquation}
\begin{split}
\frac{\dd \nN}{\dd t} = \overbrace{\lambda_{\text{N}}\nN}^{\text{proliferation}} &- \overbrace{(k_{\text{NS}}+k_{\text{NI}})\nN + k_{\text{IN}}\nI + k_{\text{SN}}\nS}^{\text{spontaneous interconversion}} + \overbrace{\delta_{\text{RA}}u_{\text{RA}}\nI}^{\text{RA-induced diff.}} - \overbrace{\delta_{\text{trk,NGF}}(1-u_{\text{NGF}})(1-u_{\text{trk}})\nN}^{\text{trkA-NGF induced apoptosis}} \\ &+ \underbrace{\delta_{\text{diff}}(1-u_{\text{trk}})u_{\text{NGF}}\nI}_{\text{trkA+NGF induced diff.}} - \underbrace{2\delta_{\text{BDNF}}(1-u_{\text{trk}})\nN.}_{\text{trkB+BDNF induced de-diff.}}
\end{split}
\end{equation}
Here, $\delta_{\text{trk,NGF}}$ is a parameter that describes the rate of apoptosis induced by trkA signaling in the absence of NGF. For the governing equation of the number of S-type cells, $\nS$, one obtains that
\begin{equation} \label{eq:nSEquation}
\begin{split}
\frac{\dd \nS}{\dd t} &= \overbrace{\lambda_{\text{S}}\nS}^{\text{proliferation}} - \overbrace{(k_{\text{SN}}+k_{\text{SI}})\nS + k_{\text{IS}}\nI + k_{\text{NS}}\nN}^{\text{spontaneous interconversion}} + \overbrace{\delta_{\text{BDNF}}(1-u_{\text{trk}})\nN.}^{\text{trkB+BDNF induced de-diff.}}
\end{split}
\end{equation}
Finally, we remark that $u_{\text{trk}}$ is an inhibitor of trk, whereas $u_{\text{RA}}, \, u_{\text{chemo}}$ and $u_{\text{NGF}}$ are all drugs that are added to the cell population. Finally, the standard bounds, ${0 \leq \mathbf{u}_k \leq 1}$, for all $k$, are imposed.

\subsection{Optimal control formulation}
\label{section:neuroblastomaOptimalControl}
The model in Equations~\eqref{eq:nIEquation}-\eqref{eq:nSEquation} contains a large number of free parameters. Each of these parameters is clinically relevant, and perhaps impossible to determine empirically. However, this complexity can be reduced in a biologically informed way, such that optimal control theory can be applied to gain insight into the effects of different parameters on treatment outcomes. This, as we will show, is inherently valuable in designing and understanding possible treatment strategies, as it allows us to identify a number of different \textit{motifs} in the shape of the pharmacodynamics profiles of the different drugs. These in turn can be used to devise novel experimental and clinical strategies. 

As a first simplification, and in the absence of knowledge of the rates of spontaneous interconversion, we choose all spontaneous interconversion constants to be equal, \textit{i.e.} ${k_{\bullet\dagger} = k \geq 0}$. As done for the example in Section~\ref{section:example1}, we rescale time, in this case by making the substitution $T = kt$ and rescaling all quantities by $k$. We furthermore assume that the effects of all the drugs on the interconversion rates can be described using a single quantity, $\delta$, \textit{i.e.}, we set $\delta_{\text{RA}} = \delta_{\text{diff}} = \delta_{\text{BDNF}} = \delta$. Finally, since the cytotoxic chemotherapeutic drug leads to cell death as a result of the cell failing to pass through the checkpoint at the end of the G2/S phase, we assume that the cell death rate is equal to $\lambda$ whenever the chemotherapeutic drug is maximally effective, \textit{i.e.}, when $u_{\text{chemo}} = 1$. To achieve this, we set $\delta_{\text{chemo}} = 2\lambda$. This results in the reduced system,
\begin{align}
    \frac{\dd \nI}{\dd t} &= (\lambda - 2)\nI + \nN+\nS - (\delta u_{\text{RA}} + 2\lambda u_{\text{chemo}} + \delta(1-u_{\text{trk}})u_{\text{NGF}})\nI + \delta(1-u_{\text{trk}})\nN, \label{eq:reducednI}\\
    \begin{split}
        \frac{\dd \nN}{\dd t} &= -2\nN + \nI + \nS + \delta(u_{\text{RA}}+(1-u_{\text{trk}})u_{\text{NGF}})\nI \\ &\phantom{(\delta_{\text{apop}}(1-u_{\text{NGF}})(1-u_{\text{trk}})} - (\delta_{\text{apop}}(1-u_{\text{NGF}})(1-u_{\text{trk}})+ 2\delta(1-u_{\text{trk}}))\nN,
    \end{split}
    \\
    \frac{\dd \nS}{\dd t} &= -2\nS + \nI + \nN + \delta(1-u_{\text{trk}})\nN, \label{eq:reducednS}
\end{align}
which only depends on three free parameters, $\delta$, $\delta_{\text{apop}}$, and $\lambda$.
The model in Equations~\eqref{eq:reducednI}-\eqref{eq:reducednS} can be cast in the form of Equation~\ref{eq:BasicEquation}, with $\mathbf{n} = (\nI, \nN, \nS)^T$, $\mathbf{u} = (u_{\text{RA}}, u_{\text{chemo}}, u_{\text{trk}}, u_{\text{NGF}})^T$, and $A$ given by
\begin{equation*}
    A = \begin{pmatrix}
    \lambda - 2 & 1 + \delta & 1\\
    1 & - 2 - 2\delta - \delta_{\text{apop}} & 1 \\
    1 & 1 + \delta & - 2
    \end{pmatrix}. 
\end{equation*}
Since there is no additive control in this model, $B = 0$. Furthermore, the terms linear in $\mathbf{u}$ are given by
\begin{equation*}
L(\mathbf{u}, \mathbf{n}) = 
    \begin{pmatrix}
        -\delta u_{\text{RA}}\nI - 2\lambda u_{\text{chemo}}\nI - 
        \delta u_{\text{NGF}}\nI
    - \delta u_{\text{trk}}\nN \\
     \delta u_{\text{RA}}\nI + 
     \delta u_{\text{NGF}}\nI +
     \delta_{\text{apop}} u_{\text{NGF}}\nN + \delta_{\text{apop}}u_{\text{trk}}\nN +  
     2\delta u_{\text{trk}}\nN \\
     -\delta u_{\text{trk}}\nN.
    \end{pmatrix}.
\end{equation*}
Therefore, the coefficient matrices, $C_\bullet$, are given by the following
\begin{equation*} C_{\text{I}} = 
    \begin{pmatrix}
        -\delta & -2\lambda& 0 & -\delta \\
        \delta& 0& 0& \delta\\
        0 & 0 & 0 & 0
    \end{pmatrix}, \quad C_{\text{N}} =
    \begin{pmatrix}
        0 & 0 & -\delta & 0 \\
        0 & 0 & \delta_{\text{apop}} + 2\delta &  \delta_{\text{apop}}\\
        0 & 0 & -\delta & 0
    \end{pmatrix},
\end{equation*}
and $C_{\text{S}} = \mathbf{0}$.

For the terms containing multi-drug interactions,
\begin{equation*}
    Q(\mathbf{u}, \mathbf{n}) = 
    \begin{pmatrix}
    \delta u_{\text{trk}}u_{\text{NGF}}\nI \\
-\delta u_{\text{trk}}u_{\text{NGF}}\nI -\delta_{\text{apop}}u_{\text{trk}}u_{\text{NGF}}\nN \\
0
    \end{pmatrix}.
\end{equation*}
Therefore, the coefficient matrices $D^{ij}$ are given by
\begin{equation*}
    D^{\text{IN}} = 
    \begin{pmatrix}
        0 & 0 & 0 &0\\
        0 & 0 & 0 & 0\\
        0 & 0 & 0 & -\delta\\
        0 & 0 & 0 & 0
    \end{pmatrix}, \quad D^{\text{NN}} = 
    \begin{pmatrix}
        0 & 0 & 0 &0\\
        0 & 0 & 0 & 0\\
        0 & 0 & 0 & -\delta_{\text{apop}}\\
        0 & 0 & 0 & 0
    \end{pmatrix},
\end{equation*}
with $D^{\text{II}} = -D^{\text{IN}}$, and all other interaction coefficient matrices satisfying $D^{ij} = \mathbf{0}$. Finally, as in Example 1, we set $R = 10^{-1} \mathbb{I}$.

\subsection{Results}
\label{section:neuroblastomaInsights}
By focusing on the reduced model in Equations~\eqref{eq:nIEquation}-\eqref{eq:nSEquation}, we can investigate the impact of varying the model parameters on the resulting drug application regimes. For all simulations we solve the model in Equations~\eqref{eq:nIEquation}-\eqref{eq:nSEquation} using the numerical scheme described in Section~\ref{section:numericalImplementation}.  At this point, we expect there to be a delicate interplay between different model parameters and the resulting optimal controls. For this reason, the first question we are interested in is what parameters have the most impact on the total dose given of each of the different drugs. To this end, we introduce the notion of maximum marginal drug cost, $c$. Given two parameters, $\xi$ and $\zeta$, we define this cost as
\begin{equation*}
    c(\zeta, \xi) = \max_{\rho} \tilde{c}(\zeta, \xi, \rho),
\end{equation*}
where $\tilde{c}$ is the optimal control cost associated with the parameter set $(\zeta, \xi, \rho)$, and $\rho$ is the parameter that is being \textit{marginalized} out. The maximum marginal cost provides a measure of which parameter regimes can incur potentially the highest drug cost. For all parameters we use the ranges $\lambda, \delta_{\text{apop}}, \delta \in [0.05, 0.5]$.

\begin{figure}[htb]
    \centering
    \includegraphics[width=\linewidth]{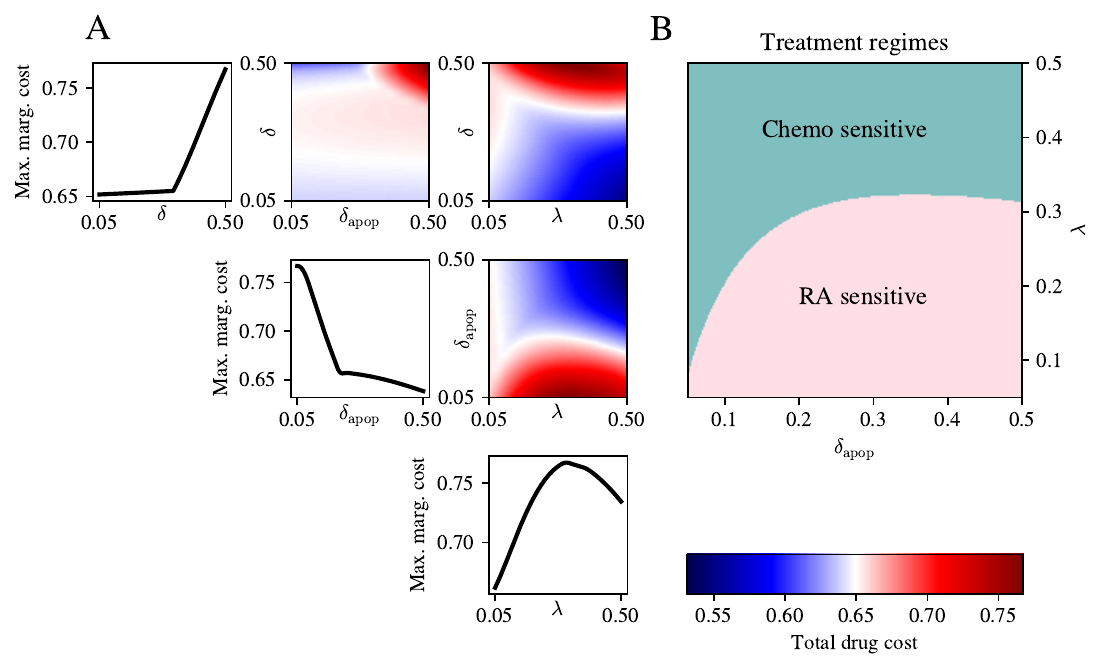}
    \caption{Effect of varying key system parameters on drug cost. A: Diagonal plots showing maximum marginal cost of drug treatment as system parameters are varied. Off-diagonal plots showing pairwise maximum marginal costs as system parameters are varied. B: Phase plane showing regions of parameter space where variation in $\delta$ leads to most change in drug response. The two sensitive drugs are chemotherapy and RA, and they form two distinct regions of the ($\delta_{\text{apop}}$, $\lambda$) space.}
    \label{fig:example2Profiles}
\end{figure}

The plots of marginal maximum drug cost in Figure~\ref{fig:example2Profiles} (panel A) show that there are clear regions of parameter space that correspond to high drug cost. The one-dimensional traces on the diagonal show that drug cost increases as the spontaneous rate of cell inter-conversions increases (as the possibly transient I-type cell population needs to be controlled constantly), decreases as the effectiveness of the trkA-NGF in inducing apoptosis increases (as then less RA is needed to induce drug kill with little NGF), and increases as the proliferation rate increases (as more chemotherapeutic agent is required to kill the cell population). As there is a clear monotonic relationship between the rate of spontaneous cell type interconversion, $\delta$, and total drug cost, we explored which drugs are most sensitive to the magnitude of this parameter, and how this is affected by the proliferation rate and the apoptosis rate. To do this, we compute a similar metric to that of maximum marginal drug cost, now intended to capture how much the concentration of each drug varies when $\delta$ is increased. We do this by computing the variance of the drug $\mathbf{u}_\bullet$, denoted $v_\bullet(\lambda, \delta_{\text{apop}})$, which we define as 
\begin{equation*}
    v_\bullet(\tilde{\lambda}, \tilde{\delta}_{\text{apop}}) = \text{Var}_{\delta}\left(\frac{1}{T}\int_0^T\mathbf{u}_\bullet\dd\tau\vert \lambda = \tilde{\lambda}, \delta_{\text{apop}} = \tilde{\delta}_{\text{apop}}\right),
\end{equation*}
that is, we capture the variability of the mean amount of drug $\mathbf{u}_\bullet$ for fixed $\lambda$ and $\delta_{\text{apop}}$ as $\delta$ is varied. We then compute for which drug this quantity is highest to produce a phase space between $\lambda$ and $\delta_{\text{apop}}$. The phase space in Figure~\ref{fig:example2Profiles} shows that there is a clear distinction in parameter space, with higher sensitivity for RA for smaller proliferation rates, $\lambda$, and higher sensitivity for chemotherapy for higher proliferation rates.  This first investigation therefore showed that by carefully inspecting how optimal controls change across parameter space, one can learn how different drugs respond to changes in parameters, which directly correspond to the biological context. This allows one to understand the expected change in dosing necessary to respond to patient variability. For example, the findings in Figure~\ref{fig:example2Profiles} suggest that in patients with rapidly dividing cancer cells (large values of $\lambda$), the amount of chemotherapy given needs to be carefully titrated according to an estimate of cell type interconversion, whereas patients with genetic features that make neuronal cells highly responsive to trkA/NGF deprivation need careful consideration of their RA dosing as RA treatment is most sensitive in that corresponding parameter regime (high $\delta_{\text{apop}}$ in Figure~\ref{fig:example2Profiles}). 

Having understood that variations in one key parameter, $\delta$, in the model in Equations~\eqref{eq:nIEquation}-\eqref{eq:nSEquation} has an unequal impact on the total amount of RA and chemotherapy that will be given according to the optimal control in different parameter regimes, we are interested in learning whether the solutions in those parameter regimes exhibit qualitative similarities in the shape of the pharmacodynamics profiles, which we call \textit{motifs}. To this end, we choose representative values in either regime and vary $\delta$ to investigate the behaviours of the model. For the RA sensitive regime, we choose $\delta_{\text{apop}} = 0.3$, $r = 0.2$. For the chemotherapy sensitive regime we choose $\delta_{\text{apop}} = 0.3$ and $r = 0.4$. In both regimes we vary the value of the cell type interconversion rate, $\delta = 0.05, 0.25, 0.45$. 

\begin{figure}
    \centering
    \includegraphics[width=\linewidth]{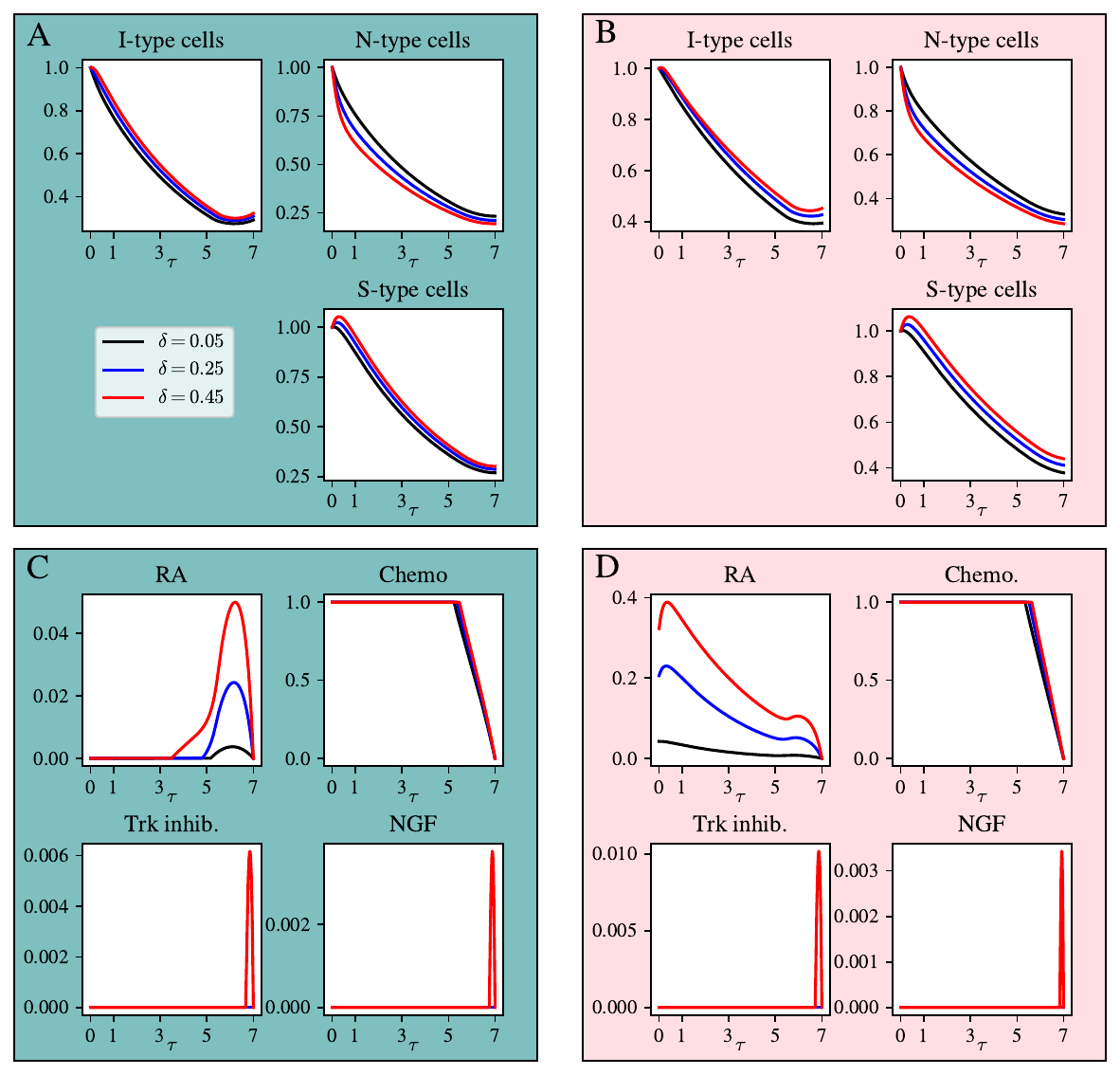}
    \caption{Optimal control of the three-population model in the chemotherapy-sensitive parameter regime and RA-sensitive regime. A: model solutions in the chemotherapy-sensitive regime, $\delta_{\text{apop}} = 0.3$ and $r = 0.4$ (teal boxes) with different values of $\delta$. B: model solutions in the RA-sensitive regime, $\delta_a = 0.3$, $r = 0.2$. C-D: corresponding optimal controls. Please note the difference in the scale of the vertical axes across the drug plots.}
    \label{fig:example2Regimes}
\end{figure}

Figure~\ref{fig:example2Regimes} shows that the pharmacodynamics profiles in the two different parameter regimes are qualitatively similar, with variations resulting from changes in the parameter $\delta$. In the chemo-sensitive regime (teal in Figures~\ref{fig:example2Profiles} and \ref{fig:example2Regimes}), there is only a small amount of RA given at the end of the treatment window, whereas in the RA-sensitive regime, all solutions start with a higher value for the pharmacodynamics at the start of the treatment window, which then decrease. For that reason, the two different parameter regimes can be interpreted as corresponding to different \textit{motifs} in pharmacodynamics. These different \textit{motifs} can be used to suggest possible distinct treatment mechanisms across different contexts. Quantitatively, varying $\delta$ has a noticeable effect on the RA dynamics in the different regimes. Put together, the analysis of this model has shown that different parameter values, corresponding to the biological characteristics of the cell population, may have a profound effect on the temporal administration of the different drugs. In particular, the complicated parameter set of this model allows to express the parameter space in regions in which the drugs are more sensitive to changes of one parameter than in other regions. 

\section{Example 3: controlling proportions of cell populations}
\label{section:example3}
Having understood how an optimal control framework can be applied to control the numbers of cells in different sub-populations, we show in this example how the framework introduced can be modified to control not the absolute cell counts of the sub-populations, but their \textit{proportions} of the total cell population. Such problems might arise, for example, in bio-engineering, when one would like to produce many clonal copies of several cell types for high-throughput experiments. To extend our framework from cell counts to cell proportions, we consider the total cell population at time $t$, 
\begin{equation*}
    N(t) = \mathbf{1}_n^T\mathbf{x}(t),
\end{equation*} 
and let $\mathbf{r}(t) = \mathbf{x}(t)/N(t)$, so that $\mathbf{r}(t)$ represents the proportion of different cell types at time $t$. Now, we have that
\begin{equation*}
    \dot{\mathbf{r}} = \frac{\dot{N}\mathbf{x}}{N^2} - \frac{N\dot{\mathbf{x}}}{N} = \frac{\dot{N}\mathbf{r} - \dot{\mathbf{x}}}{N},
\end{equation*}
and 
\begin{equation*}
    \frac{\dot{N}}{N} = \mathbf{1}_n^T\frac{\dot{\mathbf{x}}}{N},
\end{equation*}
such that 
\begin{equation}
    \label{eq:rdotFirst}
    \dot{\mathbf{r}} = \left(\mathbf{1}_n^T\frac{\dot{\mathbf{x}}}{N}\right)\mathbf{r} - \frac{\dot{\mathbf{x}}}{N}.
\end{equation}
Now, we will assume that $B =0$ in Equation~\eqref{eq:BasicEquation} and note that
\begin{equation}
    \label{eq:dotXoverN}
    \frac{\dot{\mathbf{x}}}{N} =  A\mathbf{r} + \sum_{i=1}^n C_i \odot (\mathbf{1}_n \mathbf{r}^T \mathcal{E}_i)\mathbf{u} + \sum_{i=1}^n \sum_{j=1}^n (\mathbf{e}_j^n \mathbf{r}^T\mathcal{E}_i)(D^{ij} \odot \mathbf{u}\mathbf{1}_m^T)\mathbf{u} = f(\mathbf{r}, \mathbf{u}),
\end{equation}
\textit{i.e.}, $\dot{\mathbf{r}}$ in Equation~\eqref{eq:rdotFirst} only depends on $\mathbf{r}$ and $\mathbf{u}$. This means that the optimal control framework introduced above can be used to control the proportions of cells in a population, when one is interested in this rather than their absolute cell counts. Now, one may define a target proportion, $\tilde{\mathbf{r}}$, and modify the cost function so that we penalize deviations from this target, \textit{i.e.}, 
\begin{equation*}
    J = \int_0^T \left[(\mathbf{r} - \tilde{\mathbf{r}})^T R (\mathbf{r} - \tilde{\mathbf{r}}) + \mathbf{u}^T Q \mathbf{u}\right]\dd t.
\end{equation*}
In this case, one can compute the adjoint and optimality conditions associated with this optimal control problem in a similar fashion to the previous problems. The adjoint equation (see Appendix~\eqref{appendix:proportionAdjoint}) is given by
\begin{equation*}
\begin{split}
    \dot{\boldsymbol{\lambda}} = -\frac{\partial H}{\partial \mathbf{r}} = -Q\mathbf{r} &+ (Q^T + Q)\tilde{\mathbf{r}} - (\mathbf{1}_N^T f(\mathbf{u},\mathbf{r}))\boldsymbol{\lambda} \\
    &- \left(A^T \mathbf{1}_n + \sum_{i=1}^n  (\mathbf{1}_n^TC_i\mathbf{u})\mathbf{e}_i^n + \sum_{i=1}^n \mathbf{e}_i^n \mathbf{1}_n^T\sum_{j=1}^n \mathbf{e}_j^n\mathbf{1}_m^T(D_{ij}\odot \mathbf{u}\mathbf{1}_m)\mathbf{u}\right)\odot\boldsymbol{\lambda}\odot\mathbf{r}\\
    &- \left(A^T\boldsymbol{\lambda} + \sum_{i=1}^n \mathbf{e}_i^n (C_i\mathbf{u})^T \boldsymbol{\lambda} + \sum_{i=1}^n \mathbf{e}_i \left(\sum_{j=1}^n \mathbf{e}_j^n\mathbf{1}_m^T (D^{ij}\odot \mathbf{u}\mathbf{1}_m)\mathbf{u}\right)^T\boldsymbol{\lambda}\right).
\end{split}
\end{equation*}
For the optimality condition, one finds (see Appendix~\ref{appendix:proportionOptimal}) that
\begin{equation*}
    \frac{\partial H}{\partial\mathbf{u}} = R\mathbf{u} + (\boldsymbol{\lambda}^T\mathbf{r}-1)\left(\sum_{i=1}^n\sum_{j=1}^n (\mathbf{e}_i^n)^T \mathbf{r}\boldsymbol{\lambda}^T \mathbf{e}_j^n D^{ij}\right)\mathbf{u} + \left(\sum_{i=1}^n C_i \odot (\mathbf{1}_n \mathbf{r}^T \mathcal{E}_i)\right)^T \left((\boldsymbol{\lambda}^T\mathbf{r})\mathbf{1}_n - \boldsymbol{\lambda}\right).
\end{equation*}
While the resulting equations are significantly less tractable than the optimal control conditions derived in examples 1 and 2, we note that the governing equations for the control are still given in the form of a coupled system of ODEs between $\boldsymbol{\lambda}$ and $\mathbf{r}$, meaning that they can be solved with a straightforward boundary value numerical solver. Put together, we have shown how the framework used in the previous two examples can be easily modified to account for controlling cell proportions in a population whose size can be arbitrary.

\section{Discussion}
In this work, we have presented a framework for optimally controlling the treatment response of heterogeneous populations to combination treatment. This framework describes cells of different cell types in terms of different well-mixed compartments, which interact with the administered drugs according an ODE model based on the law of mass action. As such, the framework presents a straightforward way to model an arbitrary number of cell types together with an arbitrary number of drugs. By expressing a general optimal control formulation, our framework is widely applicable, and simple to implement numerically using a boundary value problem solver. We have applied our work to three illustrative examples in mathematical biology: modeling the population of cervical cancer cells in different stages of the cell cycle when treated with two different drugs; modeling cells in different stages of development in neuroblastoma; and a general problem where one is interested in controlling the proportions of different cells in a well-mixed population.

In the system comprised of cervical cancer cells in different stages of the cell cycle, we  carefully considered the question of whether optimal control approaches can identify treatment schedules with varying pharmacodynamics that can out-perform constant dosing. We found that optimal control of two antagonistic drugs, cisplatin and paclitaxel, provides model predictions that would lead to effective treatment response than treatment with a constant action of the drug. This example is illustrative of how to one might link the notions of mathematical optimality to practical outcomes. Our approach shows that, with respect to clinically observable parameters, dosing regimes that are predicted by optimal control theory do not always share a one-to-one correspondence with dosing regimes that maximize or minimize such observables. However, our analysis shows that in certain biological contexts, optimal control theory can be used to predict regimes of administration that outperform constant dosing with respect to clinically observable parameters. The general insight from this is that one must be careful in interpreting what it means for solutions to be optimal, on the one hand, and for performing a thorough analysis of their practical utility, on the other. Of most practical interest, we found that when the total cost incurred by the drugs was low, \textit{i.e.}, when otherwise the total administered constant dose would have been small, the (time-varying) optimal control pharmacodynamics traces outperformed constant dosing. This is a promising insight therapeutically, since optimal control can provide ways to optimally design treatments at low dosage, which can help reduce the burden of drug toxicity, while improving treatment response. We also found that varying the penalty for total drug administration has a dramatic effect on the treatment profile given. While this offers tremendous opportunities to implement bespoke treatments based on patient toleration of the different drugs and differences in drug toxicity more general, it remains an open challenge to interpret and calibrate the parameters to penalize the different drug regimes.

In the example studying neuroblastoma heterogeneity, we similarly found that there is significant variability in optimal treatments depending on patient-specific parameters, such as the cell proliferation rate and the sensitivity to differentiation-inducing agents such as RA. Our approach showed that the parameter space can be studied systematically to reveal that there exist specific regions of parameter space that have characteristic shapes of the optimal control pharmacodynamics traces, which can be used to inspire dosing regimes in practice. In addition, the shapes of such pharmacodynamics traces can be linked to specific regions of parameter space, such that they might be used to probe different general themes of treatment strategies depending on the individual characteristics of the cell population under investigation. For example, we found that in some regimes, application of RA toward the end of the treatment window was most effective, whereas in others one should start with a high dose of RA. Models such as these, when calibrated to data, can be used to inform optimal treatments in a personalized manner, since individual treatment responses to individual drugs can be assessed more carefully. This, down the line, can inform treatment choices regarding which drugs to administer and how to titrate them during treatment.

There are several ways in which our framework can be extended and improved. First, our model was designed to incorporate only the pharmacodynamics of the administered drugs. In applications of optimal control to drug dosing, the question of interest becomes how the optimal solutions found through our framework might be achieved by optimal drug dosing. This would entail that the optimal control framework in this paper is be coupled to a model describing the relevant pharmacokinetics of the system under investigation. We expect this problem to be challenging computationally, due to feedback effects, \textit{e.g.} metabolism of the drug by the targeted cells. Second, the framework developed in this paper assumes that cell populations are well mixed and that cell heterogeneity is not spatial. In some contexts, however, the spatial location within a tissue or tumor is known to have an important influence on gene expression and cell identity \cite{Hausser20}. For this reason, ideas about optimal control like those explored in this paper can be extended to PDE models which directly incorporate spatial location within the tumor and the relevant anatomy to make interactions between cells and their environment more realistic. This would allow to formulate treatment plans dependent on the population environment, in addition to the population-intrinsic properties considered in this work.

As a final note, we highlight that to make advances the models presented in this work can be calibrated to experimental data. To make advances in the analysis of the models, we made assumptions about the magnitude about certain parameters, and made a careful parameter sweep to understand the behavior of the system as the parameters were varied. However, experimental methods exist to calibrate the key parameters of interest in the models to data, which can then be used to compute optimal more biologically feasible solutions for the system under investigation.

\newpage
\subsubsection*{Code availability}
Code to perform all numerical computations in this paper is accessible at our GitHub repository: \url{https://github.com/SWSJChCh/multiplicativeControl}.
\subsubsection*{Authors' contributions}
S.M.-P. conceived the project. S.M.-P., R.M.C., and S.J. developed the model in example 1. S.M.-P., R.E.B., J.K., and P.C.K. developed the model in example 2. S.M-P. and S.J. performed the analysis for the optimal control in examples 1 and 2. S.J. wrote the computational scripts used for data analysis and carried out the main analyses. S.M.-P. produced the figures and wrote the article, on which all other authors commented. All authors agreed to the publication of this manuscript.

\subsubsection*{Competing interests}
We declare we have no competing interests.

\subsubsection*{Acknowledgements}
S.M.P. would like to thank the Foulkes Foundation for funding. S.J. receives funding from the Biotechnology and Biological Sciences Research Council (BBSRC) (BB/T008784/1). R.M.C. would like to thank the Engineering and Physical Sciences Research Council (EP/T517811/1) and the Oxford-Wolfson-Marriott scholarship at Wolfson College, University of Oxford for funding. R.E.B. is supported by a grant from the Simons Foundation (MP-SIP-00001828). The authors would like to acknowledge the use of the University of Oxford Advanced Research Computing (ARC) facility in carrying out this work. 


\newpage
\bibliography{references, refs2}

\begin{thebibliography}{10}

\bibitem{lieftink2021takes}
Cor Lieftink and Roderick~L. Beijersbergen.
\newblock It takes two to tango, and the right music: Synergistic drug combinations with cell-cycle phase-dependent sensitivities.
\newblock {\em EBioMedicine}, 69, 2021.

\bibitem{Marjanovich13}
Nemanja~D. Marjanovic, Robert~A. Weinberg, and Christine~L. Chaffer.
\newblock {Cell plasticity and heterogeneity in cancer}.
\newblock {\em Clinical Chemistry}, 59(1):168--179, 2013.

\bibitem{Altschuler10}
Steven~J. Altschuler and Lani~F. Wu.
\newblock {Cellular heterogeneity: do differences make a difference?}
\newblock {\em Cell}, 141(4):559--563, 2010.

\bibitem{Hausser20}
Jean Hausser and Uri Alon.
\newblock {Tumour heterogeneity and the evolutionary trade-offs of cancer}.
\newblock {\em Nature Reviews Cancer}, 20(4):247--257, 4 2020.

\bibitem{Crucitta22}
Stefania Crucitta, Federico Cucchiara, Ron Mathijssen, Joaquin Mateo, Agnes Jager, Arjen Joosse, Antonio Passaro, Ilaria Attili, Iacopo Petrini, Ron van Schaik, Romano Danesi, and Marzia Del~Re.
\newblock {Treatment-driven tumour heterogeneity and drug resistance: Lessons from solid tumours}.
\newblock {\em Cancer Treatment Reviews}, 104, 3 2022.

\bibitem{Proietto23}
Marco Proietto, Martina Crippa, Chiara Damiani, Valentina Pasquale, Elena Sacco, Marco Vanoni, and Mara Gilardi.
\newblock {Tumor heterogeneity: preclinical models, emerging technologies, and future applications}.
\newblock {\em Frontiers in Oncology}, 13, 2023.

\bibitem{Dagogo18}
Ibiayi Dagogo-Jack and Alice~T. Shaw.
\newblock {Tumour heterogeneity and resistance to cancer therapies}.
\newblock {\em Nature Reviews Clinical Oncology}, 15(2):81--94, 2 2018.

\bibitem{Gomez22}
Roshna~Lawrence Gomez, Shakhzada Ibragimova, Revathy Ramachandran, Anna Philpott, and Fahad~R. Ali.
\newblock {Tumoral heterogeneity in neuroblastoma}.
\newblock {\em Biochimica et Biophysica Acta - Reviews on Cancer}, 1877(6), 11 2022.

\bibitem{Benninger22}
Richard~K.P. Benninger and Vira Kravets.
\newblock {The physiological role of {$\beta$}-cell heterogeneity in pancreatic islet function}.
\newblock {\em Nature Reviews Endocrinology}, 18(1):9--22, 2022.

\bibitem{Ramachandran20}
Prakash Ramachandran, Kylie~P. Matchett, Ross Dobie, John~R. Wilson-Kanamori, and Neil~C. Henderson.
\newblock {Single-cell technologies in hepatology: new insights into liver biology and disease pathogenesis}.
\newblock {\em Nature Reviews Gastroenterology and Hepatology}, 17(8):457--472, 2020.

\bibitem{lenhart07optimalcontrol}
Suzanne Lenhart and John~T. Workman.
\newblock {\em {Optimal Control Applied to Biological Models}}.
\newblock Chapman {\&} Hall, Boca Raton, 2007.

\bibitem{Pillai23}
Maalavika Pillai, Emilia Hojel, Mohit~Kumar Jolly, and Yogesh Goyal.
\newblock {Unraveling non-genetic heterogeneity in cancer with dynamical models and computational tools}.
\newblock {\em Nature Computational Science}, 3(4):301--313, 2023.

\bibitem{Clairambault23}
Jean Clairambault.
\newblock {Mathematical modelling of cancer growth and drug treatments: taking into account cell population heterogeneity and plasticity}.
\newblock In {\em 2023 European Control Conference (ECC)}, pages 1--6. EUCA, 2023.

\bibitem{Kashkooli20}
Farshad Moradi~Kashkooli, M.~Soltani, and Mohammad~Hossein Hamedi.
\newblock {Drug delivery to solid tumors with heterogeneous microvascular networks: Novel insights from image-based numerical modeling}.
\newblock {\em European Journal of Pharmaceutical Sciences}, 151, 8 2020.

\bibitem{Kirschner96}
Denise Kirschner and GF~Webb.
\newblock {A model for treatment strategy in the chemotherapy of AIDS}.
\newblock {\em Bulletin of Mathematical Biology}, 58(2):367--390, 1996.

\bibitem{Wang16}
Shuo Wang and Heinz Sch{\"{a}}ttler.
\newblock {Optimal control of a mathematical model for cancer chemotherapy under tumor heterogeneity}.
\newblock {\em Mathematical Biosciences and Engineering}, 13(6):1223--1240, 2016.

\bibitem{markman2002paclitaxel}
Maurie Markman and Tarek~M Mekhail.
\newblock Paclitaxel in cancer therapy.
\newblock {\em Expert opinion on pharmacotherapy}, 3(6):755--766, 2002.

\bibitem{Yuan16}
Lei Yuan, Gary~C. Chan, David Beeler, Lauren Janes, Katherine~C. Spokes, Harita Dharaneeswaran, Anahita Mojiri, William~J. Adams, Tracey Sciuto, Guillermo Garcia-Carde{\~{n}}a, Grietje Molema, Peter~M. Kang, Nadia Jahroudi, Philip~A. Marsden, Ann Dvorak, Erzsébet~Ravasz Regan, and William~C. Aird.
\newblock {A role of stochastic phenotype switching in generating mosaic endothelial cell heterogeneity}.
\newblock {\em Nature Communications}, 7, 2016.

\bibitem{Zeineldin22awry}
Maged Zeineldin, Anand~G. Patel, and Michael~A. Dyer.
\newblock {Neuroblastoma: When differentiation goes awry}.
\newblock {\em Neuron}, 110(18):2916--2928, 2022.

\bibitem{Jansky21Origin}
Selina Jansky, Ashwini~Kumar Sharma, Verena K{\"{o}}rber, Andrés Quintero, Umut~H. Toprak, Elisa~M. Wecht, Moritz Gartlgruber, Alessandro Greco, Elad Chomsky, Thomas~G.P. Gr{\"{u}}newald, Kai~Oliver Henrich, Amos Tanay, Carl Herrmann, Thomas H{\"{o}}fer, and Frank Westermann.
\newblock {Single-cell transcriptomic analyses provide insights into the developmental origins of neuroblastoma}.
\newblock {\em Nature Genetics}, 53(5):683--693, 5 2021.

\bibitem{Safaei12}
Farshad R.~Pour Safaei, J.~P. Hespanha, and S.~R. Proulx.
\newblock {Infinite horizon linear quadratic gene regulation in fluctuating environments}.
\newblock In {\em 2012 IEEE 51st IEEE Conference on Decision and Control (CDC)}, pages 2298--2303. IEEE, 2012.

\bibitem{shah2001cell}
Manish~A Shah and Gary~K Schwartz.
\newblock Cell cycle-mediated drug resistance: an emerging concept in cancer therapy.
\newblock {\em Clinical Cancer Research}, 7(8):2168--2181, 2001.

\bibitem{romani2022cisplatin}
Andrea Romani.
\newblock Cisplatin in cancer treatment.
\newblock {\em Biochemical pharmacology}, 206:115323, 2022.

\bibitem{muggia2000phase}
Franco~M Muggia, Patricia~S Braly, Mark~F Brady, Gregory Sutton, Theodore~H Niemann, Samuel~L Lentz, Ronald~D Alvarez, Paul~R Kucera, and James~M Small.
\newblock Phase iii randomized study of cisplatin versus paclitaxel versus cisplatin and paclitaxel in patients with suboptimal stage iii or iv ovarian cancer: a gynecologic oncology group study.
\newblock {\em Journal of Clinical Oncology}, 18(1):106--106, 2000.

\bibitem{virtanen20scipy}
Pauli Virtanen, Travis~E. Oliphant, Matt Haberland, Tyler Reddy, David Cournapeau, Evgeni Burovski, Pearu Peterson, Warren Weckesser, Stéfan~J. van~der Walt, Matthew Brett, Joshua Wilson, K.~Jarrod Millman, Nikolay Mayorov, Andrew~R.J. Nelson, Eric Jones, Robert Kern, Eric Larson, C.~J. Carey, İlhan Polat, Yu~Feng, Eric~W. Moore, Jake VanderPlas, Denis Laxalde, Josef Perktold, Robert Cimrman, Ian Henriksen, E.~A. Quintero, Charles~R. Harris, Anne~M. Archibald, Antônio~H. Ribeiro, Fabian Pedregosa, Paul van Mulbregt, Aditya Vijaykumar, Alessandro~Pietro Bardelli, Alex Rothberg, Andreas Hilboll, Andreas Kloeckner, Anthony Scopatz, Antony Lee, Ariel Rokem, C.~Nathan Woods, Chad Fulton, Charles Masson, Christian H{\"{a}}ggstr{\"{o}}m, Clark Fitzgerald, David~A. Nicholson, David~R. Hagen, Dmitrii~V. Pasechnik, Emanuele Olivetti, Eric Martin, Eric Wieser, Fabrice Silva, Felix Lenders, Florian Wilhelm, G.~Young, Gavin~A. Price, Gert~Ludwig Ingold, Gregory~E. Allen, Gregory~R. Lee, Hervé Audren, Irvin Probst,
  Jörg~P. Dietrich, Jacob Silterra, James~T. Webber, Janko Slavi{\v{c}}, Joel Nothman, Johannes Buchner, Johannes Kulick, Johannes~L. Sch{\"{o}}nberger, José~Vinícius de~Miranda~Cardoso, Joscha Reimer, Joseph Harrington, Juan Luis~Cano Rodr{\'{i}}guez, Juan Nunez-Iglesias, Justin Kuczynski, Kevin Tritz, Martin Thoma, Matthew Newville, Matthias K{\"{u}}mmerer, Maximilian Bolingbroke, Michael Tartre, Mikhail Pak, Nathaniel~J. Smith, Nikolai Nowaczyk, Nikolay Shebanov, Oleksandr Pavlyk, Per~A. Brodtkorb, Perry Lee, Robert~T. McGibbon, Roman Feldbauer, Sam Lewis, Sam Tygier, Scott Sievert, Sebastiano Vigna, Stefan Peterson, Surhud More, Tadeusz Pudlik, Takuya Oshima, Thomas~J. Pingel, Thomas~P. Robitaille, Thomas Spura, Thouis~R. Jones, Tim Cera, Tim Leslie, Tiziano Zito, and Tom Krauss.
\newblock {SciPy 1.0: fundamental algorithms for scientific computing in Python}.
\newblock {\em Nature Methods}, 17(3):261--272, 3 2020.

\bibitem{Cohn09}
Susan~L. Cohn, Andrew~D.J. Pearson, Wendy~B. London, Tom Monclair, Peter~F. Ambros, Garrett~M. Brodeur, Andreas Faldum, Barbara Hero, Tomoko Iehara, David Machin, Veronique Mosseri, Thorsten Simon, Alberto Garaventa, Victoria Castel, and Katherine~K. Matthay.
\newblock {The International Neuroblastoma Risk Group (INRG) classification system: An INRG task force report}.
\newblock {\em Journal of Clinical Oncology}, 27(2):289--297, 1 2009.

\bibitem{Zhou23Resistance}
Xia Zhou, Xiaokang Wang, Nan Li, Yu~Guo, Xiaolin Yang, and Yuhe Lei.
\newblock {Therapy resistance in neuroblastoma: Mechanisms and reversal strategies}.
\newblock {\em Frontiers in Pharmacology}, 14, 2023.

\bibitem{Korber23}
Verena K{\"{o}}rber, Sabine~A. Stainczyk, Roma Kurilov, Kai~Oliver Henrich, Barbara Hero, Benedikt Brors, Frank Westermann, and Thomas H{\"{o}}fer.
\newblock {Neuroblastoma arises in early fetal development and its evolutionary duration predicts outcome}.
\newblock {\em Nature Genetics}, 55(4):619--630, 4 2023.

\bibitem{Groningen17}
Tim Van~Groningen, Jan Koster, Linda~J. Valentijn, Danny~A. Zwijnenburg, Nurdan Akogul, Nancy~E. Hasselt, Marloes Broekmans, Franciska Haneveld, Natalia~E. Nowakowska, Johannes Bras, Carel~J.M. Van~Noesel, Aldo Jongejan, Antoine~H. Van~Kampen, Linda Koster, Frank Baas, Lianne Van Dijk-Kerkhoven, Margriet Huizer-Smit, Maria~C. Lecca, Alvin Chan, Arjan Lakeman, Piet Molenaar, Richard Volckmann, Ellen~M. Westerhout, Mohamed Hamdi, Peter~G. Van~Sluis, Marli~E. Ebus, Jan~J. Molenaar, Godelieve~A. Tytgat, Bart~A. Westerman, Johan Van~Nes, and Rogier Versteeg.
\newblock {Neuroblastoma is composed of two super-enhancer-associated differentiation states}.
\newblock {\em Nature Genetics}, 49(8):1261--1266, 8 2017.

\bibitem{Maris10}
John~M Maris.
\newblock {Recent advances in neuroblastoma}.
\newblock {\em The New England Journal of Medicine}, 3060:2202--2213, 2010.

\bibitem{Schmelz21}
Karin Schmelz, Joern Toedling, Matt Huska, Maja~C. Cwikla, Louisa~Marie Kruetzfeldt, Jutta Proba, Peter~F. Ambros, Inge~M. Ambros, Sengül Boral, Marco Lodrini, Celine~Y. Chen, Martin Burkert, Dennis Guergen, Annabell Szymansky, Kathy Astrahantseff, Annette Kuenkele, Kerstin Haase, Matthias Fischer, Hedwig~E. Deubzer, Falk Hertwig, Patrick Hundsdoerfer, Anton~G. Henssen, Roland~F. Schwarz, Johannes~H. Schulte, and Angelika Eggert.
\newblock {Spatial and temporal intratumour heterogeneity has potential consequences for single biopsy-based neuroblastoma treatment decisions}.
\newblock {\em Nature Communications}, 12(1), 12 2021.

\bibitem{Brodeur14Regression}
Garrett~M. Brodeur and Rochelle Bagatell.
\newblock {Mechanisms of neuroblastoma regression}.
\newblock {\em Nature Reviews Clinical Oncology}, 11(12):704--713, 12 2014.

\bibitem{Matthay99}
Katherine~K. Matthay, Judith~G. Villablanca, Robert~C. Seeger, Daniel~O. Stram, Richard~E. Harris, Norma~K. Ramsay, Patrick Swift, Hiroyuki Shimada, C.~Thomas Black, Garrett~M. Brodeur, Robert~B. Gerbing, and C.~Patrick Reynolds.
\newblock {Treatment of high-risk neuroblastoma with intensive chemotherapy, radiotherapy, autologous bone marrow transplantation and 13-cis-retinoic acid}.
\newblock {\em The New England Journal of Medicine}, 341:1165--1173, 1999.

\bibitem{Matthay16}
Katherine~K. Matthay, John~M. Maris, Gudrun Schleiermacher, Akira Nakagawara, Crystal~L. Mackall, Lisa Diller, and William~A. Weiss.
\newblock {Neuroblastoma}.
\newblock {\em Nature Reviews Disease Primers}, 2, 11 2016.

\bibitem{Boeva17}
Valentina Boeva, Caroline Louis-Brennetot, Agathe Peltier, Simon Durand, Cécile Pierre-Eug{\`{e}}ne, Virginie Raynal, Heather~C. Etchevers, Sophie Thomas, Alban Lermine, Estelle Daudigeos-Dubus, Birgit Geoerger, Martin~F. Orth, Thomas~G.P. Gr{\"{u}}newald, Elise Diaz, Bertrand Ducos, Didier Surdez, Angel~M. Carcaboso, Irina Medvedeva, Thomas Deller, Valérie Combaret, Eve Lapouble, Gaelle Pierron, Sandrine Grosset{\^{e}}te-Lalami, Sylvain Baulande, Gudrun Schleiermacher, Emmanuel Barillot, Hermann Rohrer, Olivier Delattre, and Isabelle Janoueix-Lerosey.
\newblock {Heterogeneity of neuroblastoma cell identity defined by transcriptional circuitries}.
\newblock {\em Nature Genetics}, 49(9):1408--1413, 9 2017.

\bibitem{Kasemeier18Logic}
Jennifer~C Kasemeier-Kulesa, Santiago Schnell, Thomas Woolley, Jennifer~A Spengler, Jason~A Morrison, Mary~C McKinney, Irina Pushel, Lauren~A Wolfe, and Paul~M Kulesa.
\newblock {Predicting neuroblastoma using developmental signals and a logic-based model}.
\newblock {\em Biophysical Chemistry}, 238:30--38, 2018.

\bibitem{Kasemeier-Kulesa2015TrkB/BDNFSystem}
Jennifer~C. Kasemeier-Kulesa, Jason~A. Morrison, Frances Lefcort, and Paul~M. Kulesa.
\newblock {TrkB/BDNF signalling patterns the sympathetic nervous system}.
\newblock {\em Nature Communications}, 6, 2015.

\bibitem{Kasemeier18}
Jennifer~C. Kasemeier-Kulesa, Morgan~H. Romine, Jason~A. Morrison, Caleb~M. Bailey, Danny Rwelch, and Paul~M. Kulesa.
\newblock {NGF reprograms metastatic melanoma to a bipotent glial-melanocyte neural crest-like precursor}.
\newblock {\em Biology Open}, 7(1), 2018.

\bibitem{Iyer16}
Radhika Iyer, Lea Wehrmann, Rebecca~L Golden, Koumudi Naraparaju, Jamie~L Croucher, Suzanne~P MacFarland, Peng Guan, Venkatadri Kolla, Ge~Wei, Nicholas Cam, Gang Li, Zachary Hornby, and Garrett~M Brodeur.
\newblock {Entrectinib is a potent inhibitor of Trk-driven neuroblastomas in a xenograft mouse model}.
\newblock {\em Cancer letters}, 372(2):179--186, 2016.

\end{thebibliography}
\bibliographystyle{unsrt}

\newpage
\appendix
\setcounter{equation}{0}\renewcommand\theequation{A\arabic{equation}}
\section{Computation of the adjoint conditions}
\label{appendix:Adjoint}
We compute  the adjoint equations defined by
\begin{equation*}
    \dot{\lambda}_k = - \frac{\partial H}{\partial \mathbf{x}_k}.
\end{equation*}
Note first
\begin{equation*}
    \frac{\partial}{\partial \mathbf{x}_k} \boldsymbol{\lambda}^T \sum_{i=1}^n C_i \odot(\mathbf{1}_n \mathbf{x}^T\mathcal{E}_i)\mathbf{u} = \frac{\partial}{\partial \mathbf{x}_k} \boldsymbol{\lambda}^T C_k \odot(\mathbf{1}_n \mathbf{x}^T\mathcal{E}_k)\mathbf{u} = \frac{\partial}{\partial \mathbf{x}_k} \mathbf{x}_k \boldsymbol{\lambda}^T C_k \odot(\mathbf{1}_n \mathbf{1}_n^T\mathcal{E}_k)\mathbf{u},
\end{equation*}
such that
\begin{equation*}
    \frac{\partial}{\partial \mathbf{x}_k} \boldsymbol{\lambda}^T \sum_{i=1}^n C_i \odot(\mathbf{1}_n \mathbf{x}^T\mathcal{E}_i)\mathbf{u} = \boldsymbol{\lambda}^T C_k \mathbf{u},
\end{equation*}
where the last equality is due to the fact that $\mathbf{1}_n \mathbf{1}_n^T\mathcal{E}_k$ is an $n\times n$ matrix filled with ones. Hence
\begin{equation*}
    \frac{\partial}{\partial \mathbf{x}} \boldsymbol{\lambda}^T \sum_{i=1}^n C_i \odot(\mathbf{1}_n \mathbf{x}^T\mathcal{E}_i)\mathbf{u} = \sum_{i=1}^n \mathbf{e}_i^n \boldsymbol{\lambda}^T C_i\mathbf{u} = \left(\sum_{i=1}^n \mathbf{e}_i^n (C_i \mathbf{u})^T\right)\boldsymbol{\lambda}.
\end{equation*}
Similarly, we obtain
\begin{equation*}
    \frac{\partial}{\partial \mathbf{x}_k} \boldsymbol{\lambda}^T \sum_{i=1}^n \sum_{j=1}^n(\mathbf{e}_j^n \mathbf{x}^T\mathcal{E}_i)(D^{ij}\odot \mathbf{u}\mathbf{1}_m^{T})\mathbf{u} = \boldsymbol{\lambda}^T \sum_{j=1}^n \mathbf{e}_j^n\mathbf{1}_n\mathcal{E}_k(D^{ij}\odot \mathbf{u}\mathbf{1}_m^{T})\mathbf{u}.
\end{equation*}
Now, noticing
\begin{equation*}
    \mathbf{e}_j^n \mathbf{1}_n^T \mathcal{E}_k  = \mathbf{e}_j^n\mathbf{1}_n^T \mathbf{e}_k^n \mathbf{1}_m^T = \mathbf{e}_j^n\mathbf{1}_m^T,
\end{equation*}
we find
\begin{equation*}
    \frac{\partial}{\partial \mathbf{x}_k} \boldsymbol{\lambda}^T \sum_{i=1}^n \sum_{j=1}^n(\mathbf{e}_j^n \mathbf{x}^T\mathcal{E}_i)(D^{ij}\odot \mathbf{u}\mathbf{1}_m^{T})\mathbf{u} = \boldsymbol{\lambda}^T \sum_{j=1}^n \mathbf{e}_j^n\mathbf{1}_m^T(D^{ij}\odot \mathbf{u}\mathbf{1}_m^{T})\mathbf{u},
\end{equation*}
such that
\begin{equation*}
    \frac{\partial}{\partial \mathbf{x}} \boldsymbol{\lambda}^T \sum_{i=1}^n \sum_{j=1}^n(\mathbf{e}_j^n \mathbf{x}^T\mathcal{E}_i)(D^{ij}\odot \mathbf{u}\mathbf{1}_m^{T})\mathbf{u} = \left(\sum_{i=1}^n \mathbf{e}_i \sum_{j=1}^n \mathbf{e}_j^n\mathbf{1}_m^T(D^{ij}\odot \mathbf{u}\mathbf{1}_m^{T})\mathbf{u}\right)^T \boldsymbol{\lambda}.
\end{equation*}
Put together,
\begin{equation}
    \label{eq:finalAdjointComp}
    \frac{\partial}{\partial\mathbf{x}} \boldsymbol{\lambda}^T\dot{\mathbf{x}} =  A^T\boldsymbol{\lambda} - \sum_{i=1}^n \mathbf{e}_i^n (C_i\mathbf{u})^T \boldsymbol{\lambda} - \sum_{i=1}^n \mathbf{e}_i \left(\sum_{j=1}^n \mathbf{e}_j^n\mathbf{1}_m^T (D^{ij}\odot \mathbf{u}\mathbf{1}_m^{T})\mathbf{u}\right)^T\boldsymbol{\lambda}.
\end{equation}

\newpage
\section{Computation of the optimality conditions}
\label{appendix:Optimality}
Notice first
\begin{equation}
    \label{eq:firstOptimalityComp}
    \frac{\partial}{\partial \mathbf{u}} \boldsymbol{\lambda}^T \sum_{i=1}^n C_i \odot (\mathbf{1}_n \mathbf{x}^T \mathcal{E}_i) \mathbf{u} = \left(\sum_{i=1}^n C_i \odot (\mathbf{1}_n \mathbf{x}^T \mathcal{E}_i)\right)^T \boldsymbol{\lambda}.
\end{equation}
Now we compute
\begin{equation*}
    \frac{\partial}{\partial \mathbf{u_{k}}} \boldsymbol{\lambda}^T (\mathbf{e}_{j}^n \mathbf{x}^T \mathcal{E}_{i})(D^{ij}\odot \mathbf{u}\mathbf{1}_{m})\mathbf{u} = \frac{\partial}{\partial \mathbf{u_{k}}} \boldsymbol{\lambda}^T \mathbf{\xi^{ij}}, 
\end{equation*}
where $\mathbf{\xi^{ij}} = (\mathbf{e}_{j}^n \mathbf{x}^T \mathcal{E}_{i})(D^{ij}\odot \mathbf{u}\mathbf{1}_{m})\mathbf{u}$ is an $n\times1$ vector with a non-zero element $\alpha_{ij}$ only in the $j^{\text{th}}$ entry:
\begin{equation}
\label{sam1}
    \alpha_{ij} = \sum_{p=1}^{m}u_{p}\sum_{q=1}^{m}x_{i}D^{ij}_{qp}u_{q},
\end{equation}
and $D^{ij}$ is a matrix as defined previously. In differentiating $\mathbf{\xi^{ij}}$ with respect to $\mathbf{u}_{k}$, we apply simple product rule differentiation to \eqref{sam1} to obtain
\begin{align*}
    \frac{\partial}{\partial \mathbf{u_{k}}}\alpha_{ij} &= \sum_{p\neq k}\mathbf{u}_{p}\mathbf{x}_{i}D_{qp}^{ij}\mathbf{u}_{q} + \mathbf{u}_{k}D^{ij}_{qp}\mathbf{u}_{q} + \mathbf{u}_{k}D^{ij}_{kk}\mathbf{u}_{i} \\
    &= \sum_{p = 1}^{n}\mathbf{u}_{p}\mathbf{x}_{i}D_{qp}^{ij} + \mathbf{u}_{k}D_{kk}^{ij}\mathbf{x}_i,
\end{align*}
where we have used the lower triangularity of $D^{ij}$. Now
\begin{equation*}
    \frac{\partial}{\partial \mathbf{u}_k}\boldsymbol{\lambda}^T (\mathbf{e}_j^n \mathbf{x}^T\mathcal{E}_i)(D^{ij}\odot \mathbf{u}\mathbf{1}_m)\mathbf{u} = \boldsymbol{\lambda} \left[\sum_{p=1}^n \mathbf{u}_p\mathbf{x}_i (D^{ij})_{kp} + \mathbf{u_k}(D^{ij})_{kk}\mathbf{x}_i\right].
\end{equation*}
Defining now the matrix $\mathcal{D}^k$ by
\begin{equation*}
    (\mathcal{D}^k)_{ij} = (D^{ij})_{kk},
\end{equation*}
we find
\begin{equation*}
    \sum_{i=1}^n\sum_{j=1}^n \mathbf{u}_k (D^{ij})_{kk} \mathbf{x}_i \boldsymbol{\lambda}_j = \mathbf{u}_k \mathbf{1}^T \mathcal{D}^k \odot (\mathbf{x}\boldsymbol{\lambda}^T)\mathbf{1},
\end{equation*}
and also
\begin{equation*}
    \sum_{i=1}^n \sum_{j=1}^n \mathbf{x}_i \boldsymbol{\lambda}_j (D^{ij})\mathbf{u}\cdot \mathbf{e}_k = \sum_{i=1}^n \sum_{j=1}^n (\mathbf{e}_i^n)^T \mathbf{x}\boldsymbol{\lambda}^T\mathbf{e}_j^n (D^{ij}\mathbf{u})\cdot \mathbf{e}_k^m,
\end{equation*}
and hence
\begin{equation*}
    \frac{\partial}{\partial \mathbf{u}_k} \boldsymbol{\lambda}^T \sum_{i=1}^n \sum_{j=1}^n (\mathbf{e}_j^n \mathbf{x}^T \mathcal{E}_i)(D^{ij} \odot \mathbf{u} \mathbf{1}_m)\mathbf{u} = \mathbf{u}_k \mathbf{1}^T_m \mathcal{D}^k\odot(\mathbf{x}\boldsymbol{\lambda}^T)\mathbf{1}_m + \sum_{i=1}^n \sum_{j=1}^n (\mathbf{e}_i^n)^T \mathbf{x}\boldsymbol{\lambda}^T\mathbf{e}_j^n (D^{ij}\mathbf{u})\cdot \mathbf{e}_k^m,
\end{equation*}
which gives
\begin{equation}
    \label{eq:secondOptimalitycomp}
    \frac{\partial}{\partial \mathbf{u}} \boldsymbol{\lambda}^T \sum_{i=1}^n \sum_{j=1}^n (\mathbf{e}_j^n \mathbf{x}^T \mathcal{E}_i)(D^{ij} \odot \mathbf{u} \mathbf{1}_m)\mathbf{u} = \left(\sum_{i=1}^m \mathbf{e}_i \mathbf{1}^T \mathcal{D}^i \odot (\mathbf{x}\boldsymbol{\lambda}^T)\mathbf{1}\right)\odot \mathbf{u} + \left(\sum_{i=1}^n\sum_{j=1}^n (\mathbf{e}_i^n)^T \mathbf{x}\boldsymbol{\lambda}^T \mathbf{e}_j^n D^{ij}\right)\mathbf{u}.
\end{equation}
Finally, note that for the model we have assumed that $\mathcal{D}^i = 0$ for all $i$ since there are no terms in the form $\mathbf{u}_k\mathbf{u}_k$ for any $k$ in any of the equations. For that reason,
\begin{equation*}
    \frac{\partial}{\partial \mathbf{u}} \boldsymbol{\lambda}^T \sum_{i=1}^n \sum_{j=1}^n (\mathbf{e}_j^n \mathbf{x}^T \mathcal{E}_i)(D^{ij} \odot \mathbf{u} \mathbf{1}_m)\mathbf{u} =  \left(\sum_{i=1}^n\sum_{j=1}^n (\mathbf{e}_i^n)^T \mathbf{x}\boldsymbol{\lambda}^T \mathbf{e}_j^n D^{ij}\right)\mathbf{u}.
\end{equation*}
Now putting everything together, we obtain an expression for $\partial H/\partial \mathbf{u}$ given by
\begin{equation*}
    \frac{\partial H}{\partial \mathbf{u}} = R\mathbf{u} + B^T\boldsymbol{\lambda}  +  \left(\sum_{i=1}^n C_i \odot (\mathbf{1}_n \mathbf{x}^T \mathcal{E}_i)\right)^T \boldsymbol{\lambda}  + \left(\sum_{i=1}^n\sum_{j=1}^n \mathbf{e}_i^n \mathbf{x}\boldsymbol{\lambda}^T \mathbf{e}_j^n D^{ij}\right)\mathbf{u}.
\end{equation*}
\newpage
\section{Solution of the optimality condition}
Imposing the restriction $\mathbf{u}_k \in [0,1]$ for all $k$ implies via the standard theory of optimal control that 
\begin{equation*}
    \mathbf{u}^*_k = \begin{cases} 
      0 & \frac{\partial H}{\partial \mathbf{u}_k} < 0, \\
      0 \leq \mathbf{u}_k \leq 1 & \frac{\partial H}{\partial \mathbf{u}_k} = 0,\\
      1 & \frac{\partial H}{\partial \mathbf{u}_k} > 0. 
   \end{cases}
\end{equation*}
Concretely, the middle case corresponds to solutions of the equation
\begin{equation*}
    \mathbf{0} = R\mathbf{u} + B^T\boldsymbol{\lambda}  +  \left(\sum_{i=1}^n C_i \odot (\mathbf{1}_n \mathbf{x}^T \mathcal{E}_i)\right)^T \boldsymbol{\lambda}  + \left(\sum_{i=1}^n\sum_{j=1}^n \mathbf{e}_i^n \mathbf{x}\boldsymbol{\lambda}^T \mathbf{e}_j^n D^{ij}\right)\mathbf{u},
\end{equation*}
if they exist. In that case, the optimality condition reads
\begin{equation*}
    \left(R+\sum_{i=1}^n\sum_{j=1}^n \mathbf{e}_i^n \mathbf{x}\boldsymbol{\lambda}^T \mathbf{e}_j^n D^{ij}\right)\mathbf{u} = -\left(B+\sum_{i=1}^n C_i \odot (\mathbf{1}_n \mathbf{x}^T \mathcal{E}_i)\right)^T\boldsymbol{\lambda},
\end{equation*}
and so, if the inverse exists,
\begin{equation*}
    \mathbf{u}^{\star} = -\left(R+\sum_{i=1}^n\sum_{j=1}^n \mathbf{e}_i^n \mathbf{x}\boldsymbol{\lambda}^T \mathbf{e}_j^n D^{ij}\right)^{-1}\left(B+\sum_{i=1}^n C_i \odot (\mathbf{1}_n \mathbf{x}^T \mathcal{E}_i)\right)^T\boldsymbol{\lambda}.
\end{equation*}

\newpage
\section{On why there is no closed-form expression for a feedback control in Example 1}
\label{appendix:RicattiEquation}
In the case of the LQR problem, which has dynamics given by Equation~\eqref{eq:LQR} and optimizes the cost function given in Equation~\eqref{eq:costFunction}, it can be shown that the optimal control is given by 
\begin{equation*}
    \mathbf{u}^* = -R^{-1} B^T S \mathbf{x},
\end{equation*}
where $S$ is a matrix that satisfies a quadratic first-order ODE called the Ricatti equation \cite{lenhart07optimalcontrol}.  In optimal control terminology, the matrix $R^{-1}B^TS$ is called the \textit{gain}. The matrix $S$ arises from making the ansatz $\mathbf{\lambda}(t) = S(t)\mathbf{x}(t)$. Making this substitution, it follows that
\begin{equation*}
    \dot{\boldsymbol{\lambda}} = \dot{S}(t)\mathbf{x} + S(t)\dot{\mathbf{x}}(t),
\end{equation*}
so that, remembering that $(D^{ij})_{qp} = 0$ for all $p \geq q$,
\begin{equation*}
    \dot{S}(t)\mathbf{x} + S(t)\dot{\mathbf{x}}(t) = \dot{\boldsymbol{\lambda}} = -Q\mathbf{x} - A^T\boldsymbol{\lambda} - \sum_{i=1}^n \mathbf{e}_i^n (C_i\mathbf{u})^T \boldsymbol{\lambda},
\end{equation*}
from which we obtain that
\begin{align*}
    \dot{S}(t)\mathbf{x} &=-S(t)\dot{\mathbf{x}}(t) -Q\mathbf{x} - A^T\boldsymbol{\lambda} - \sum_{i=1}^n \mathbf{e}_i^n (C_i\mathbf{u})^T \boldsymbol{\lambda}\\
    &= -S(t)\left(A\mathbf{x} + \sum_{i=1}^n C_i \odot (\mathbf{1}_n \mathbf{x}^T \mathcal{E}_i)\mathbf{u} \right) -Q\mathbf{x} - A^T\boldsymbol{\lambda} - \sum_{i=1}^n \mathbf{e}_i^n (C_i\mathbf{u})^T \boldsymbol{\lambda}.
\end{align*}
At this point, if $\mathbf{u}$ were unbounded, one would substitute the control, Equation~\eqref{eq:ReducedOptimalControl}, to compute
\begin{align*}
    \sum_{i=1}^n C_i \odot (\mathbf{1}_n \mathbf{x}^T \mathcal{E}_i)\mathbf{u} &= \sum_{i=1}^n C_i \odot (\mathbf{1}_n \mathbf{x}^T \mathcal{E}_i)\left(-R^{-1}\left(\sum_{j=1}^n C_j \odot (\mathbf{1}_n \mathbf{x}^T \mathcal{E}_j)\right)^T\boldsymbol{\lambda}\right)\\
    &= - \left(\sum_{i=1}^n \sum_{j=1}^n [C_i \odot (\mathbf{1}_n \mathbf{x}^T \mathcal{E}_i)] R^{-1} [C_j \odot (\mathbf{1}_n \mathbf{x}^T \mathcal{E}_j)]^T\right)\boldsymbol{\lambda}\\
    &= - \left(\sum_{i=1}^n \sum_{j=1}^n [C_i \odot (\mathbf{1}_n \mathbf{x}^T \mathcal{E}_i)] R^{-1} [C_j \odot (\mathbf{1}_n \mathbf{x}^T \mathcal{E}_j)]^T\right)S\mathbf{x}\\
    &\equiv \mathcal{R}_1 \mathbf{x}.
\end{align*}
As for the second term involving $\mathbf{u}$, we compute
\begin{align*}
    \sum_{i=1}^n \mathbf{e}_i^n (C_i\mathbf{u})^T \boldsymbol{\lambda}&= -\sum_{i=1}^n \mathbf{e}_i^n \left(C_iR^{-1}\left(\sum_{j=1}^n C_j \odot (\mathbf{1}_n \mathbf{x}^T \mathcal{E}_j)\right)^T\boldsymbol{\lambda}\right)^T \boldsymbol{\lambda}\\
    &= -\sum_{i=1}^n \mathbf{e}_i^n \left(\boldsymbol{\lambda}^T \left(\sum_{j=1}^n C_j \odot (\mathbf{1}_n \mathbf{x}^T \mathcal{E}_j)\right) R^{-T}C_i^T\right) \boldsymbol{\lambda}\\
    &= -\sum_{i=1}^n \mathbf{e}_i^n \left(\mathbf{x}^TS \left(\sum_{j=1}^n C_j \odot (\mathbf{1}_n \mathbf{x}^T \mathcal{E}_j)\right) R^{-T}C_i^T\right) S\mathbf{x}\\\
    &\equiv \mathcal{R}_2 \mathbf{x}.
\end{align*}
Put together, we would obtain
\begin{equation*}
    \dot{S}\mathbf{x} = \left[-SA + \mathcal{R}_1 - Q - A^TS - \mathcal{R}_2\right]\mathbf{x},
\end{equation*}
so that
\begin{equation}
    \label{eq:fakeRicatti}
    \dot{S} = \left[-SA + \mathcal{R}_1 - Q - A^TS - \mathcal{R}_2\right] = M.
\end{equation}
Notice that Equation~\eqref{eq:fakeRicatti} is strictly speaking not a Ricatti equation as it is linear in $S$. Proceeding one would substitute back to find that the control is only a type of \textit{feedback control},
\begin{equation*}
    \mathbf{u}^{\star} = -R^{-1}\left(\sum_{i=1}^n C_i \odot (\mathbf{1}_n \mathbf{x}^T \mathcal{E}_i)\right)^TS\mathbf{x}.
\end{equation*}
The complication in this procedure arises as a result of the bounds imposed on $\mathbf{u}$. In the substitution of the  control given by Equation~\eqref{eq:ReducedOptimalControl}, no bounds were assumed. Therefore, the closed form solution for the equation for $\dot{S}$ that arises takes no restrictions on $\mathbf{u}$ into account. We note that in numerical implementations, one can easily incorporate these bounds when computing the derivative of $S$.

\newpage
\section{Computation of optimality and adjoint equations for Example 3 in the main text}
\subsection{Adjoint conditions}
\label{appendix:proportionAdjoint}
We first compute the quantity $\partial H/\partial \mathbf{r}$. Since 
\begin{equation*}
    (\mathbf{r} - \tilde{\mathbf{r}})^T Q (\mathbf{r} - \tilde{\mathbf{r}}) = \mathbf{r}^T Q\mathbf{r} + \tilde{\mathbf{r}}^T Q \tilde{\mathbf{r}} - (Q^T\tilde{\mathbf{r}})\cdot \mathbf{r} - (Q\tilde{\mathbf{r}})\cdot\mathbf{r},
\end{equation*}
we find that
\begin{equation*}
    \frac{\partial H}{\partial \mathbf{r}} = Q\mathbf{r} - (Q^T + Q)\tilde{\mathbf{r}} + \frac{\partial}{\partial \mathbf{r}} \boldsymbol{\lambda}^T \dot{\mathbf{r}}.
\end{equation*}
The third term is given by
\begin{align*}
        \frac{\partial}{\partial \mathbf{r}_k} \boldsymbol{\lambda}^T \dot{\mathbf{r}} &= \frac{\partial}{\partial \mathbf{r}_k}\boldsymbol{\lambda}^T \left(\left(\mathbf{1}_n^T\frac{\dot{\mathbf{x}}}{N}\right)\mathbf{r} - \frac{\dot{\mathbf{x}}}{N}\right)\\
        &= \boldsymbol{\lambda}_k\left(\mathbf{1}_n^T\frac{\dot{\mathbf{x}}}{N}\right) + \boldsymbol{\lambda}_k\mathbf{r}_k\left(\frac{\partial}{\partial \mathbf{r}_k} \mathbf{1}_n^T \frac{\dot{\mathbf{x}}}{N}\right) - \frac{\partial}{\partial \mathbf{r}_k}\left(\boldsymbol{\lambda}^T\frac{\dot{\mathbf{x}}}{N}\right).
\end{align*}
Observe first that the final term in the last expression is equal to the $k^{\text{th}}$ component of the expression in Equation~\eqref{eq:finalAdjointComp} with $\mathbf{x}$ replaced by $\mathbf{r}$. Now, the following computation is essentially the same as that for the adjoint condition that we did previously, but we include it for completeness here
\begin{align*}
    \frac{\partial}{\partial\mathbf{r}_k}\left(\mathbf{1}_n^T \frac{\dot{\mathbf{x}}}{N}\right) &= \frac{\partial}{\partial\mathbf{r}_k}\mathbf{1}_n^T\left( A\mathbf{r} + \sum_{i=1}^n C_i \odot (\mathbf{1}_n \mathbf{r}^T \mathcal{E}_i)\mathbf{u} + \sum_{i=1}^n \sum_{j=1}^n (\mathbf{e}_j^n \mathbf{r}^T\mathcal{E}_i)(D^{ij} \odot \mathbf{u}\mathbf{1}_m^T)\mathbf{u} \right) \\
    &= \mathbf{1}_n^T A\mathbf{e}_k^n + \frac{\partial}{\partial \mathbf{r}_k}\left[\mathbf{1}_n^T\left(C_k\odot(\mathbf{1}_n\mathbf{r}^T\mathcal{E}_i)\mathbf{u}\right) + \mathbf{1}_n^T\sum_{j=1}^n (\mathbf{e}_j\mathbf{r}^T\mathcal{E}_k)(D_{kj}\odot\mathbf{u}\mathbf{1}_m)\mathbf{u} \right],
\end{align*}
where the second equality is due to the fact that the matrix $C_i\odot(\mathbf{1}_n^T\mathbf{r}^T\mathcal{E}_i)$ contains only terms in $\mathbf{r}_i$ for each $i$, and the matrix $(\mathbf{e}_j^n \mathbf{r}^T\mathcal{E}_i)$ contains only $\mathbf{r}_i$ on the $j$-th row for each $0 \leq i,j \leq n$. Proceeding, and using the fact that the aforementioned matrices are constant in $\mathbf{r}_i$ for each $i$,
\begin{align*}
    \frac{\partial}{\partial\mathbf{r}_k}\left(\mathbf{1}_n^T \frac{\dot{\mathbf{x}}}{N}\right) &= \mathbf{1}_n^T A\mathbf{e}_k^n + \frac{\partial}{\partial \mathbf{r}_k}\left[\mathbf{r}_k\mathbf{1}_n^T\left(C_k\odot(\mathbf{1}_n\mathbf{1}_n^T\mathcal{E}_k)\mathbf{u}\right) + \mathbf{r}_k\mathbf{1}_n^T\sum_{j=1}^n (\mathbf{e}_j\mathbf{1}_n^T\mathcal{E}_k)(D_{kj}\odot\mathbf{u}\mathbf{1}_m)\mathbf{u} \right]\\
    &= \mathbf{1}_n^T A\mathbf{e}_k^n +\mathbf{1}_n^T\left(C_k\odot(\mathbf{1}_n\mathbf{1}_n^T\mathcal{E}_k)\mathbf{u}\right) + \mathbf{1}_n^T\sum_{j=1}^n (\mathbf{e}_j\mathbf{1}_n^T\mathcal{E}_k)(D_{kj}\odot\mathbf{u}\mathbf{1}_m)\mathbf{u}, 
\end{align*}
and so we conclude that
\begin{align*}
    \frac{\partial}{\partial \mathbf{r}} \left(\mathbf{1}_n^T \frac{\dot{\mathbf{x}}}{N}\right) &= A\mathbf{1}_n +\sum_{i=1}^n\mathbf{e}_i^n \mathbf{1}_n^T\left(C_i\odot(\mathbf{1}_n\mathbf{1}_n^T\mathcal{E}_i)\mathbf{u}\right) + \sum_{i=1}^n \mathbf{e}_i^n \mathbf{1}_n^T\sum_{j=1}^n (\mathbf{e}_j\mathbf{1}_n^T\mathcal{E}_k)(D_{kj}\odot\mathbf{u}\mathbf{1}_m)\mathbf{u}\\
    &= A \mathbf{1}_n + \sum_{i=1}^n  (\mathbf{1}_n^TC_i\mathbf{u})\mathbf{e}_i^n + \sum_{i=1}^n \mathbf{e}_i^n \mathbf{1}_n^T\sum_{j=1}^n \mathbf{e}_j^n\mathbf{1}_m^T(D_{ij}\odot \mathbf{u}\mathbf{1}_m)\mathbf{u}.
\end{align*}
Letting $f(\mathbf{u}, \mathbf{r})$ be as in Equation~\eqref{eq:dotXoverN}, we obtain
\begin{equation*}
\begin{split}
    \frac{\partial H}{\partial \mathbf{r}} = Q\mathbf{r} &- (Q^T + Q)\tilde{\mathbf{r}} + (\mathbf{1}_N^T f(\mathbf{u},\mathbf{r}))\boldsymbol{\lambda} \\
    &+ \left(A \mathbf{1}_n + \sum_{i=1}^n  (\mathbf{1}_n^TC_i\mathbf{u})\mathbf{e}_i^n + \sum_{i=1}^n \mathbf{e}_i^n \mathbf{1}_n^T\sum_{j=1}^n \mathbf{e}_j^n\mathbf{1}_m^T(D_{ij}\odot \mathbf{u}\mathbf{1}_m)\mathbf{u}\right)\odot\boldsymbol{\lambda}\odot\mathbf{r}\\
    &+ \left(A^T\boldsymbol{\lambda} + \sum_{i=1}^n \mathbf{e}_i^n (C_i\mathbf{u})^T \boldsymbol{\lambda} + \sum_{i=1}^n \mathbf{e}_i \left(\sum_{j=1}^n \mathbf{e}_j^n\mathbf{1}_m^T (D^{ij}\odot \mathbf{u}\mathbf{1}_m)\mathbf{u}\right)^T\boldsymbol{\lambda}\right).
\end{split}
\end{equation*}

\subsection{Optimality conditions}
\label{appendix:proportionOptimal}
For the optimality condition, we compute
\begin{equation*}
    \frac{\partial H}{\partial \mathbf{u}} = R\mathbf{u} + \frac{\partial}{\partial \mathbf{u}} \boldsymbol{\lambda}^T\dot{\mathbf{r}} = R\mathbf{u} + \frac{\partial}{\partial \mathbf{u}} \boldsymbol{\lambda}^T\left(\left(\mathbf{1}_n^T\frac{\dot{\mathbf{x}}}{N}\right)\mathbf{r} - \frac{\dot{\mathbf{x}}}{N}\right).
\end{equation*}
Notice that, as in Appendix~\ref{appendix:proportionAdjoint}, the term given by $(\partial/\partial\mathbf{u}) \left(\boldsymbol{\lambda}^T \frac{\dot{\mathbf{x}}}{N}\right)$ is equal to that given in the optimality condition for $\mathbf{x}$ with $\mathbf{x}$ replaced by $\mathbf{r}$. Therefore, we only need to compute the first term. Notice that
\begin{equation*}
    \frac{\partial}{\partial \mathbf{u}_k}\boldsymbol{\lambda}^T\left(\mathbf{1}_n^T\frac{\dot{\mathbf{x}}}{N}\right)\mathbf{r} = \frac{\partial}{\partial \mathbf{u}_k} \left(\mathbf{1}_n^T \frac{\dot{\mathbf{x}}}{N}\right)\boldsymbol{\lambda}^T\mathbf{r}, 
\end{equation*}
and
\begin{align*}
    \frac{\partial}{\partial \mathbf{u}_k} \left(\mathbf{1}_n^T \frac{\dot{\mathbf{x}}}{N}\right) &= \frac{\partial}{\partial \mathbf{u}_k} \boldsymbol{1}_n^T\left(A\mathbf{r} + \sum_{i=1}^n C_i \odot (\mathbf{1}_n \mathbf{r}^T \mathcal{E}_i)\mathbf{u} + \sum_{i=1}^n \sum_{j=1}^n (\mathbf{e}_j^n \mathbf{r}^T\mathcal{E}_i)(D^{ij} \odot \mathbf{u}\mathbf{1}_m^T)\mathbf{u}\right) \\
    &= \frac{\partial}{\partial \mathbf{u}_k} \boldsymbol{1}_n^T\left(\sum_{i=1}^n C_i \odot (\mathbf{1}_n \mathbf{r}^T \mathcal{E}_i)\mathbf{u} + \sum_{i=1}^n \sum_{j=1}^n (\mathbf{e}_j^n \mathbf{r}^T\mathcal{E}_i)(D^{ij} \odot \mathbf{u}\mathbf{1}_m^T)\mathbf{u} \right). \\
\end{align*}
From Equation~\eqref{eq:firstOptimalityComp}, one can deduce that
\begin{equation*}
    \frac{\partial}{\partial \mathbf{u}} \boldsymbol{1}_n^T\left(\sum_{i=1}^n C_i \odot (\mathbf{1}_n \mathbf{r}^T \mathcal{E}_i)\mathbf{u}\right) = \left(\sum_{i=1}^n C_i \odot (\mathbf{1}_n \mathbf{r}^T \mathcal{E}_i)\right)^T \mathbf{1}_n.
\end{equation*}
Likewise, from Equation~\eqref{eq:secondOptimalitycomp}, one obtains that 
\begin{equation*}
    \frac{\partial}{\partial \mathbf{u}} \mathbf{1}_n^T\left( \sum_{i=1}^n \sum_{j=1}^n (\mathbf{e}_j^n \mathbf{r}^T\mathcal{E}_i)(D^{ij} \odot \mathbf{u}\mathbf{1}_m^T)\mathbf{u} \right) = \left(\sum_{i=1}^n\sum_{j=1}^n (\mathbf{e}_i^n)^T \mathbf{r}\mathbf{1}_n^T \mathbf{e}_j^n D^{ij}\right)\mathbf{u}.
\end{equation*}
Put together,
\begin{equation*}
\begin{split}
        \frac{\partial H}{\partial \mathbf{u}} = R\mathbf{u} &+ (\boldsymbol{\lambda}^T\mathbf{r})\left(\left(\sum_{i=1}^n C_i \odot (\mathbf{1}_n \mathbf{r}^T \mathcal{E}_i)\right)^T \mathbf{1}_n + \left(\sum_{i=1}^n\sum_{j=1}^n (\mathbf{e}_i^n)^T \mathbf{r}\mathbf{1}_n^T \mathbf{e}_j^n D^{ij}\right)\mathbf{u}\right)\\
        &- \left(\sum_{i=1}^n C_i \odot (\mathbf{1}_n \mathbf{x}^T \mathcal{E}_i)\right)^T \boldsymbol{\lambda} -\left(\sum_{i=1}^n\sum_{j=1}^n (\mathbf{e}_i^n)^T \mathbf{r}\boldsymbol{\lambda}^T \mathbf{e}_j^n D^{ij}\right)\mathbf{u}.
\end{split}
\end{equation*}
Re-writing this expression now yields
\begin{equation*}
        \frac{\partial H}{\partial\mathbf{u}} = R\mathbf{u} + (\boldsymbol{\lambda}^T\mathbf{r}-1)\left(\sum_{i=1}^n\sum_{j=1}^n (\mathbf{e}_i^n)^T \mathbf{r}\boldsymbol{\lambda}^T \mathbf{e}_j^n D^{ij}\right)\mathbf{u} + \left(\sum_{i=1}^n C_i \odot (\mathbf{1}_n \mathbf{r}^T \mathcal{E}_i)\right)^T \left((\boldsymbol{\lambda}^T\mathbf{r})\mathbf{1}_n - \boldsymbol{\lambda}\right).
\end{equation*}


\end{document}